\documentclass{emulateapj}
\usepackage{apjfonts}
\newcommand{\cmden}{\mbox{ cm$^{-3}$}}
\newcommand{\colden}{\mbox{ cm$^{-2}$}}

\newcommand{\ergs}{\mbox{ erg s$^{-1}$}}
\newcommand{\kms}{\mbox{ km s$^{-1}$}}
\newcommand{\emiss}{\mbox{ erg cm$^{-3}$ s$^{-1}$}}

\newcommand{\Mpcden}{\mbox{ Mpc$^{-3}$}}

\newcommand{\kel}{\mbox{ K}}
\newcommand{\Mpc}{\mbox{ Mpc}}
\newcommand{\kpc}{\mbox{ kpc}}

\newcommand{\msun}{\mbox{ M$_\odot$}}
\newcommand{\Zsun}{Z_\odot}
\newcommand{\sfr}{\mbox{ M$_\odot$ yr$^{-1}$}}
\newcommand{\secinv}{\mbox{ s$^{-1}$}}

\newcommand{\junits}{\mbox{ erg s$^{-1}$ cm$^{-2}$ Hz$^{-1}$ sr$^{-1}$}}
\newcommand{\photunits}{\mbox{ photons s$^{-1}$ cm$^{-2}$ sr$^{-1}$}}
\newcommand{\photfluxA}{\mbox{ photons cm$^{-2}$ s$^{-1}$ sr$^{-1}$
    \AA$^{-1}$}}
\newcommand{\sberg}{\mbox{ erg s$^{-1}$ cm$^{-2}$ arcsec$^{-2}$}}

\newcommand{\hunits}{\mbox{ km s$^{-1}$ Mpc$^{-1}$}}
\newcommand{\bq}{\begin{equation}}
\newcommand{\eq}{\end{equation}}
\newcommand{\lya}{Ly$\alpha$ }
\newcommand{\nh}{n_{\rm H}}

\newcommand{\epsa}{\epsilon_\alpha}

\begin{document}

\title{Ly$\alpha$ Emission from Structure Formation}

\author{Steven R. Furlanetto\altaffilmark{1}, Joop
Schaye\altaffilmark{2}, Volker Springel\altaffilmark{3}, \& Lars
Hernquist\altaffilmark{4}}

\altaffiltext{1} {Mail Code 130-33; California Institute of
  Technology; Pasadena, CA 91125; sfurlane@tapir.caltech.edu}

\altaffiltext{2} {School of Natural Sciences, Institute for Advanced
Study, Einstein Drive, Princeton NJ 08540}

\altaffiltext{3} {Max-Planck-Institut f\"{u}r Astrophysik,
Karl-Schwarzschild-Strasse 1, 85740 Garching, Germany}

\altaffiltext{4} {Harvard-Smithsonian Center for Astrophysics, 60
Garden St., Cambridge, MA 02138}

\begin{abstract}

The nature of the interaction between galaxies and the intergalactic
medium (IGM) is one of the most fundamental problems in astrophysics.
The accretion of gas onto galaxies provides fuel for star formation,
while galactic winds transform the nearby IGM in a number of ways.
One exciting technique to study this gas is through the imaging of
hydrogen \lya emission.  We use cosmological simulations to study the
\lya signals expected from the growth of cosmic structure from
$z=0$--$5$.  We show that if dust absorption is negligible,
recombinations following the absorption of stellar ionizing photons
dominate the total \lya photon production rate.  However, galaxies are
also surrounded by ``\lya coronae'' of diffuse IGM gas.  These coronae
are composed of a combination of accreting gas and material ejected
from the central galaxy by winds.  The \lya emission from this phase
is powered by a combination of gravitational processes and the
photoionizing background.  While the former dominates at $z \sim 0$,
collisional excitation following photo-heating may well dominate the
total emission at higher redshifts.  The central regions of these
systems are dense enough to shield themselves from the metagalactic
ionizing background; unfortunately, in this regime our simulations are
no longer reliable.  We therefore consider several scenarios for the
emission from the central cores, including one in which self-shielded
gas does not emit at all.  We show that the combination of star
formation and cooling IGM gas can explain most of the observed ``\lya
blobs'' at $z \sim 3$, with the important exception of the largest
sources.  On the other hand, except under the most optimistic
assumptions, cooling IGM gas cannot explain the observations on its
own.

\end{abstract}


\section{Introduction}
\label{intro}

The distribution of matter on large scales is one of the key
ingredients of any cosmological paradigm.  A number of observations
constraining the overall cosmological paradigm, together with
numerical simulations, have shown that the ``cosmic web'' provides a
good description of the Universe on the largest scales.  In this model,
structure formation proceeds through the collapse of matter into
sheets, filaments, and eventually galaxies and galaxy clusters.
However, on small scales, baryons and dark matter are subject to
different forces, and the resulting complications have severely
limited the success of ab initio models in predicting the structure
and evolution of galaxies. The interactions between galaxies and the
intergalactic medium (IGM) are particularly fundamental to our
understanding of the formation and growth of galaxies. The accretion
of gas onto galaxies provides fuel for star formation, while feedback
processes (both radiative and mechanical) from star formation can
transform the IGM.

Quasar absorption lines are the most well-known way to study the IGM
at moderate redshifts, and there is no doubt that the \lya forest has
provided a wealth of insight into the IGM, the distribution of gas
around galaxies, and the nature of the interactions between the two
(see \citealt{rauch98} for a review). An important aspect of the
forest is that absorbers are localized in redshift so that each can be
easily separated in the redshift direction.  However, because the
forest probes only isolated lines of sight, it has proved difficult to
determine how the absorption systems relate to the cosmic web and to
galaxies.  These are particularly intriguing questions because the
answers will shed light on the formation and growth of galaxies.  An
alternate approach is to search for \lya \emph{emission} from hydrogen
in galaxies and in the IGM. This method has the advantage of
permitting a direct reconstruction of the three-dimensional gas
distribution, which will allow much cleaner constraints on the
interplay with galaxies.  Moreover, the \lya radiation directly probes
an important fraction of the cooling radiation at the temperatures
relevant for the formation of galaxies, and it is often more readily
detected than continuum radiation.  For these reasons, this approach
has received a large amount of attention, despite the challenging
nature of the observations.

In particular, narrowband Ly$\alpha$-selected surveys have emerged in
recent years as an efficient way to identify and localize
high-redshift objects.  Such surveys have been used to detect faint
galaxy associations at moderate redshifts (\citealt{steidel00};
hereafter S00), high redshift galaxies
\citep{hue02,kodaira03,rhoads04}, the filamentary structure of
galaxies at high redshifts \citep{moller01}, the host galaxies of
Lyman-limit and damped \lya systems (e.g., \citealt{fynbo99,fynbo00}),
and diffuse emission near radio galaxies \citep{mccarthy87}.  Ongoing
efforts to detect \lya emission from high-column density absorbers in
the IGM have already yielded interesting constraints \citep{francis04}
and will soon reach the predictions of simple models \citep{gould96}.

The last effort is an example of one of the most exciting
possibilities for \lya surveys: imaging gas \emph{outside} of
galaxies.  To date, most Ly$\alpha$-selected sources are compact
star-forming galaxies.  Ionizing photons from young, massive stars are
absorbed by the interstellar media of the galaxies and re-radiated
(following recombinations) as \lya photons.  However, the ubiquity of
hydrogen also permits the direct detection of the IGM.  One example is
the diffuse \lya emission often found near radio galaxies
\citep{mccarthy87}, thought to be shock-heated gas interacting with
the radio jet.  Another, perhaps more representative, set of objects
are extended ``\lya blobs'' at $z \sim 3$ with little or no continuum
emission.  S00 detected two distinct regions of diffuse \lya
emission around (but not centered on) three of the Lyman-break
galaxies in their field (which contained 27 such galaxies).  These
objects have $L_{\rm Ly\alpha} \sim 10^{44} \ergs$ and physical
extents of $\sim 100 h^{-1} \kpc$.  More recent observations revealed
33 smaller blobs in an overlapping field, many clustered around known
galaxies (\citealt{matsuda04}; hereafter M04).  These \lya blobs are
particularly intriguing because they sit at the interface between
galaxies and the IGM and should probe the interactions between the
two.

There are a number of processes that can cause \lya emission from the
IGM, most of which involve interactions with galaxies.
\citet{hogan87} suggested that optically thick, high-column density
systems should absorb ionizing photons from the metagalactic radiation
field and re-radiate a substantial fraction of them as \lya photons.
\citet{gould96} refined this argument and showed that deep
observations with 8-meter class telescopes could detect the resulting
emission.  Interestingly, the luminosity of a system scales with the
local ionizing background in this scenario (so long as it remains
optically thick).  Thus the \lya blobs could be extreme examples of
this effect, in which ionizing photons from luminous, embedded sources
are reprocessed into \lya photons.  The largest blob does contain a
luminous submm (though optically invisible) source with an inferred
star formation rate (SFR) $\ga 500 h^{-2} \sfr$ \citep{chapman01}, which
could power the extended emission.  An alternate possibility is the
assembly of gravitationally bound objects.  In the traditional view,
baryons are shocked as they accrete onto halos \citep{white78}.  They
then shed their gravitational energy through radiative cooling, and a
large fraction of the energy could be lost as \lya radiation (e.g.,
\citealt{haiman00-lya}). \citet{fardal01} examined this mechanism with
cosmological simulations and argued that it could explain the luminous
\lya blobs, although they find weaker shocks than expected (see also
\citealt{birnboim03,keres04}).  Finally, there is evidence for strong
galactic winds near the most luminous blobs
\citep{steidel00,taniguchi01,bower04}, and the \lya emission could be
a result of the interaction of this wind with the surrounding IGM.
Similar interactions probably explain the diffuse \lya emission
surrounding radio galaxies at high redshifts \citep{mccarthy87}.

Unfortunately, these scenarios are difficult to distinguish, and there
has been relatively little theoretical investigation of the
possibilities.  In \citet[hereafter F03]{furl03-lya}, we used a
state-of-the-art cosmological simulation to predict the \lya emission
from IGM gas at $z \la 0.5$.  We found that most galaxies were
surrounded by ``coronae'' of \lya emission a few times larger than the
embedded galaxies.  We argued that the emitting gas is slowly cooling
onto the galaxies, so that the \lya emission traces the growth of
galaxies.  In this paper, we extend our work to higher redshifts $0 <
z < 5$ and examine in detail the origin of the emission and of the gas
responsible for it. We show that \lya emission can be particularly
powerful during the era at which galaxy assembly peaks ($z \sim
1$--$3$) and we discuss the relative importance of the
different physical mechanisms that are responsible for the production
of \lya photons from diffuse gas.

At the same time, it is useful to consider \lya emission from inside
of galaxies.  Massive, hot stars produce ionizing photons, most of
which are absorbed by the surrounding hydrogen gas.  As this gas
recombines, it radiates \lya photons; the \lya luminosity thus offers
a (rough) measure of the star formation rate, although dust and
radiative transfer through the galaxy and IGM can decouple the two in
individual objects (e.g., \citealt{shapley03}).  The SFR in turn
depends (at least crudely) on the rate at which gas accretes onto the
galaxy, so this mode also contains information on how galaxies grow.
The stellar ionizing photons can also escape into the IGM,
illuminating the neighborhood of each galaxy and boosting the
brightness of the \lya coronae.  We thus also use the simulations to
predict the expected emission that can be traced to stellar ionizing
photons.  We will show that the emission from the IGM and from star
formation both contribute to any given source, but that the former is
more spatially extended.

We first discuss the numerical simulations and our analysis procedure
in \S \ref{sims}.  We then describe how to compute the \lya emissivity
of a parcel of gas in \S \ref{emiss}.  We review the properties of the
IGM, and how they relate to the \lya emission, in \S \ref{phase}.  We
then describe our results in terms of the statistical properties of
maps in \S \ref{maps} and in terms of individual objects in \S
\ref{source}.  Finally, we conclude in \S \ref{disc}.

We present most of our results in terms of surface brightness.  We use
two (interchangeable) sets of units; for reference, $1 \mbox{ photon
s$^{-1}$ cm$^{-2}$ sr$^{-1}$} = 9.6 \times 10^{-23} [4/(1+z)]$ erg
s$^{-1}$ cm$^{-2}$ arcsec$^{-2}$.  The redshift factor appears because
$\lambda_{\rm obs} = \lambda_\alpha (1+z)$, where
$\lambda_\alpha=1215.67$ \AA \ is the rest wavelength of the hydrogen
\lya transition.

\section{Simulations and Analysis Procedure}
\label{sims}

We use the suite of cosmological simulations described by
\citet{springel03} and \citet{nagamine04}.  These smoothed-particle
hydrodynamics (SPH) simulations include a multiphase description of
star formation incorporating a prescription for galactic winds
\citep{springel-sf} and were performed using a fully conservative
form of SPH \citep{springel-ent}.  They also accounted for the
presence of a uniform ionizing background as described by Haardt \&
Madau (1996). The simulations assume a $\Lambda$CDM cosmology with
$\Omega_m=0.3$, $\Omega_\Lambda=0.7$, $\Omega_b=0.04$, $H_0=100 h
\hunits$ (with $h=0.7$), and a scale-invariant primordial power
spectrum with index $n=1$ normalized to $\sigma_8=0.9$ at the present
day.  These parameters are consistent with the most recent
cosmological observations (e.g., \citealt{spergel03}).  Because we
wish to study a variety of redshifts and physical scales, we will use
several different simulations.  We summarize their main parameters in
Table \ref{table:sims}.

\begin{deluxetable}{ccccccc}
  \tablehead{ \colhead{Name} & \colhead{$L$} & \colhead{Resolution} &
    \colhead{$m_{\rm gas}$} & \colhead{$\epsilon$} & \colhead{$z_{\rm
    end}$} & \colhead{Winds}} \tablecaption{Simulations} \startdata Q5
    & 10.00 & $2 \times 324^3$ & $3.26 \times 10^5$ & 1.23 & 2.75 &
    Strong \\ Q4 & 10.00 & $2 \times 216^3$ & $1.10 \times 10^6$ &
    1.85 & 2.75 & Strong \\ Q3 & 10.00 & $2 \times 144^3$ & $3.72
    \times 10^6$ & 2.78 & 2.75 & Strong \\ P3 & 10.00 & $2 \times
    144^3$ & $3.72 \times 10^6$ & 2.78 & 2.75 & Weak \\ O3 & 10.00 &
    $2 \times 144^3$ & $3.72 \times 10^6$ & 2.78 & 2.75 & None \\ D5 &
    33.75 & $2 \times 324^3$ & $1.26 \times 10^7$ & 4.17 & 1 & Strong
    \\ G6 & 100.0 & $2 \times 486^3$ & $9.67 \times 10^7$ & 5.00 & 0 &
    Strong \\ G5 & 100.0 & $2 \times 324^3$ & $3.26 \times 10^8$ &
    8.00 & 0 & Strong \enddata \tablecomments{All simulations are
    described in \citet{springel03}, with the exception of G6 (which
    is described in \citealt{nagamine04}). Here $L$ is the box size in
    $h^{-1}$ comoving Mpc and $\epsilon$ is the softening length in
    $h^{-1}$ comoving kpc.  The ``Resolution'' column gives the
    initial number of particles (including both gas and dark
    matter). The particle mass $m_{\rm gas}$ is listed in $h^{-1}
    \msun$.  ``Strong'' and ``weak'' winds have nominal velocities of
    $484$ and $242 \kms$, respectively. \label{table:sims}}
\end{deluxetable}

To make maps of the \lya emission, we wish to compute the surface
brightness of a slice of the simulation with fixed $z$, $\Delta z$,
and angular size.  \citet{furl04-metals} describe our analysis
procedure in detail, so we only summarize it here.  We first randomly
choose the volume corresponding to this slice from the appropriate
simulation output.  We then divide the projected volume into a grid of
$N^2$ pixels; in most of what follows we choose $N=512$.  For each
particle in the simulation, we check whether it lies within the slice
of interest.  If so, we compute its \lya emissivity by linearly
interpolating (in log space) the grids described in \S \ref{emiss},
assuming that the particle is cubical with a uniform density (this is
reasonable because luminous particles are much more compact than our
map pixels).  We then determine which pixel(s) the particle overlaps
and add the particle's surface brightness to them, weighting by the
fraction of the pixel subtended by the particle.  Finally, we smooth
the maps using a Gaussian filter with a width of four pixels; the
resolution $\Delta \theta$ that we quote is the FWHM of this smoothing
function.

\section{Ly$\alpha$ Emission}
\label{emiss}

Our emissivity calculations follow, for the most part, the same
procedures as in F03 and \citet{furl04-metals}.  We first divide the
gas into three regimes: optically thin, self-shielded, and
star-forming. The first component consists of gas elements that are
optically thin to the background ionizing radiation field.  We use the
photoionizing backgrounds of \citet{haardt01}, which include emission
from galaxies and quasars as well as reprocessing by the IGM.  We
denote the total ionizing rate by $\Gamma \equiv \Gamma_{12} \times
10^{-12} \secinv$.  For reference, we have $\Gamma_{12} =
(0.086,0.790,1.50,1.15,0.761,0.565)$ at $z=(0,1,2,3,4,5)$.  

\subsection{Optically Thin Gas}
\label{photo}

To compute the \lya emissivity $\epsa$ of optically thin gas, we
construct a grid in density $\nh$ with spacing $\Delta \log \nh=0.25$
in the range $-6 < \log \nh < -1$ and $\Delta \log \nh=1.0$ elsewhere
(here $\nh$ is measured in cm$^{-3}$) and another grid in temperature
with $\Delta \log T=0.05$ in the range $3.5 < \log T < 5$ and $\Delta
\log T=0.25$ elsewhere (here $T$ is measured in degrees K).  We then
use the Cloudy 96 photoionization code (beta 5; \citealt{cloudy96}) to
compute the emissivity at each of the grid points, assuming solar
metallicity.  (The results are insensitive to the metallicity, so long
as it is near or below this value.)

\begin{figure}[t]
\plotone{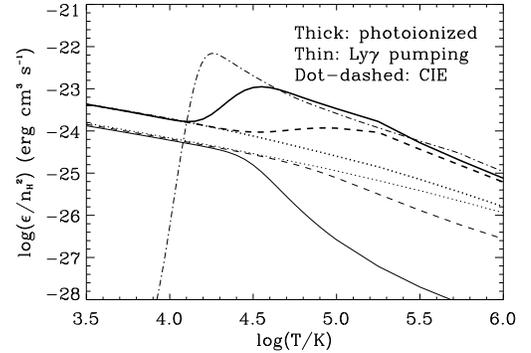}
\caption{Ly$\alpha$ emissivities at $z=3$.  The thick curves show
  $\epsilon_\alpha/\nh^2$ for $\log (\nh/{\rm cm^{-3}})=-2,\,-4$, and
  $-6$ (solid, dashed, and dotted curves, respectively).  The thin
  curves show the pumping emissivity from Ly$\gamma$, assuming an
  optically thin medium, for the same densities.  The dot-dashed curve
  shows the Ly$\alpha$ emissivity for gas in collisional ionization
  equilibrium (CIE).}
\label{fig:emiss}
\end{figure}

The thick curves in Figure~\ref{fig:emiss} show $\epsa/\nh^2$ for
three different densities at $z=3$ in the optically thin case; other
redshifts are qualitatively similar.  At small $\nh$, collisional
excitation can be ignored and $\epsa$ essentially follows the
temperature dependence of the recombination coefficient.  In this
regime, the emissivity is independent of the amplitude of the ionizing
background.  As the density increases, collisional processes become
increasingly important and cause the increase in $\epsa/\nh^2$ at $T
\sim 10^{4.5}$--$10^5 \kel$.  We note that in this regime excitation
dominates the emissivity, with recombinations (following either
collisional or photoionization) making only a small contribution.  The
emissivity does depend on $\Gamma$ if the density is large enough that
the gas is no longer highly ionized ($n_{\rm HI}/n_{\rm H} \sim 0.5
n_{\rm H} [T/10^4~{\rm K}]^{-0.76} \Gamma_{12}^{-1}$, e.g.\ Schaye
2001b).

\subsection{Self-shielded Gas}
\label{ss}

If the column density of a cloud is sufficiently high, gas in the
central regions becomes self-shielded from the metagalactic background and
ionizing radiation cannot penetrate the cloud.  Unfortunately, because
the simulations do not include radiative transfer, we cannot identify
these clouds self-consistently. Instead we note that photoionized,
self-gravitating clouds in (local) hydrostatic equilibrium obey
\citep{schaye}
\bq 
N_{\rm HI} \sim 2.3 \times 10^{13} \colden \ \left( \frac{\nh}{10^{-5}
  \cmden} \right)^{3/2} \ \left( \frac{T}{10^4 \kel} \right)^{-0.26}
\Gamma_{12}^{-1}, 
\label{eq:sscol}
\eq
where $N_{\rm HI}$ is the column density of neutral hydrogen.  (Here,
we have set the mass fraction of gas in the cloud to
$\Omega_b/\Omega_m$.)  If we assume that the gas in the simulation is
close to hydrostatic equilibrium, we can then identify self-shielded
clouds through the physical densities reported by the simulation. The
clouds should become self-shielded above some column density threshold
that depends on the shape of the ionizing background (because higher
energy photons have smaller absorption cross-sections).  To fix this
density, we follow \citet{schaye01-damp}, who performed detailed
radiative transfer through such clouds.  These calculations show that
self-shielding becomes significant at $N_{\rm HI} \sim 10^{18}
\colden$ (see also \citealt{katz96a}).  At $z=0$ we find this column
density to correspond to a physical density threshold $n_{\rm H,ss} =
10^{-3} \cmden$.  We also find $n_{\rm H,ss} \propto \Gamma^{2/3}$ in
the range of interest, as predicted by equation (\ref{eq:sscol}); the
relatively modest evolution in the spectral shape of the
\citet{haardt01} ionizing background has little effect on the result.
For simplicity, we will use this scaling to determine how $n_{\rm
H,ss}$ evolves with redshift. Of course, dense gas will not
self-shield if it is collisionally ionized. To crudely account for
this, we will assume that gas must have $T<T_{\rm ss} \equiv 10^{4.5}
\kel$ to be self-shielded.  

Now that we have identified it, we must compute $\epsa$ for shielded
gas.  In principle, this is straightforward.  Without an ionizing
background, most of the gas will be in approximate collisional
ionization equilibrium (CIE), and $\epsa$ is again simply a function
of the local density and temperature.  In this case, we compute
$\epsa$ with Cloudy through a grid over temperature; the spacing is
$\Delta \log T=0.02$ in the range $3.7 < \log T < 4.3$ and $\Delta
\log T=0.05$ elsewhere (where $T$ is in degrees K).  The recombination
rate at $T \la 10^{3.7} \kel$ depends slightly on metallicity, so we
also compute a grid in metallicity with spacing of one dex.  However,
$\epsa$ is so small in this regime that the metallicity turns out to
have only a negligible effect on our results.  We do not need a grid
in density, because $\epsa \propto \nh^2$ in CIE.  Note that, except
at the highest temperatures, \lya following collisional excitation
dominates by a large amount over \lya emission following
recombinations.  We show $\epsa$ for gas in CIE as the dot-dashed
curve in Figure~\ref{fig:emiss}.  Note that it has a much steeper
temperature dependence than optically thin gas has.  We shall see that
this has important consequences for our different treatments of
self-shielded gas.

The simplest prescription, and our first case, is then to take
particle temperatures reported by the simulation at face value and
assume CIE.  Unfortunately, this is probably not accurate.  As
\citet{fardal01} point out, the introduction of an ionizing background
into any simulation without a self-consistent treatment of radiative
transfer leads to an unphysical energy exchange between self-shielded
gas and the ionizing background.  In our simulation, the thermal
evolution of each particle is computed assuming that it is exposed to
a uniform ionizing background that photoheats the gas.  As we shall
see below, cool dense gas approaches an asymptotic temperature at
which photoheating nearly balances radiative cooling.  Although
self-shielded particles can still cool radiatively, they (by
definition) do not experience any heating from the ionizing
background.  The simulation therefore overestimates their temperature.
Using the model of \citet{schaye01-damp}, we find that if photoheating
from the metagalactic background were the \emph{only} heat source,
these particles would quickly cool below $T \sim 10^{4.1} \kel$ and
produce little or no \lya emission.  Thus our second case -- and the
most conservative -- is to set $\epsa=0$ for self-shielded gas.  Of
course, the temperature could remain high because of other heat
sources that are not included in the simulation.

On the other hand, it is possible that some fraction (or even most) of
the gas that we label ``self-shielded'' is actually photoionized.
Nearly all self-shielded regions in the simulation surround galaxies,
so the local ionizing background may exceed the mean by a large
factor.  Our third case is thus to assume that \emph{all} gas is
optically thin and to use the simulation temperatures.  (Note that the
simulation temperatures could still be wrong if the local ionizing
radiation differs from the mean background.  However, the temperature
dependence is relatively modest for optically thin gas, so such errors
are not as important in this prescription.)  Our first case, in which
we assume CIE, is in some sense between the second and third: the
local ionizing background is strong enough to heat the gas, but weak
enough that CIE is a good approximation.  Obviously, none of these
solutions is either elegant or fully self-consistent, but we expect
that they should bracket reality.

We note that \citet{fardal01} took a different approach and used a
simulation without an ionizing background.  In this case the particle
temperatures are fixed exclusively by gravitational processes and
galactic feedback.  \emph{If} photoheating can be neglected before the
gas becomes self-shielded, this would yield more accurate temperatures
for the dense, shielded gas.  However, we will show below that
photoheating is usually important for the \lya emission.

Our treatment of cool, dense gas has a number of additional
uncertainties.  First, in computing its emissivity, we assume that the
gas is optically thin to Lyman line radiation.  This is obviously not
true; in fact, many photons from the higher Lyman lines will be
absorbed inside the cloud and downgraded to either a single \lya
photon or a pair of 2$s$ continuum photons through radiative cascades.
This could increase the \lya emissivity by a factor of a few in the
best cases.  Unfortunately, the fraction of these photons that are
reprocessed into \lya photons depends on the column depth of the
cloud, its geometry, and its velocity structure (because photons
escape when they scatter into the wings of the Lyman lines).  Given
the other uncertainties, we do not attempt to model this
amplification.  Second, we neglect absorption by dust in the cloud;
this is particularly important if the \lya photons scatter many times
before escaping.  Our simulations do follow the metallicity of each
gas particle, so we could in principle estimate this effect.  But it
too depends on the geometry and velocity structure of the cloud (e.g.,
\citealt{neufeld91}), and we will not attempt to do so.
Because the IGM metallicity is normally modest, we expect dust
absorption to be less important in these environments than inside
galaxies. 

\subsection{Star-forming Particles}
\label{sflum}

So far we have considered gas outside of galaxies.  Active star
formation (SF) also produces substantial Ly$\alpha$ emission when gas
in the host galaxy absorbs ionizing radiation from stars and
subsequently recombines.  With a slight abuse of terminology, we will
refer to this mechanism as the SF component.  In the simulations, SF
occurs in any gas particle whose density exceeds $\nh=0.129 \cmden$
\citep{springel-sf}.  This threshold was fixed through comparison to
observations of star-forming galaxies \citep{kennicutt89,kennicutt98},
which show an analogous surface density threshold for star
formation. Independent models of self-gravitating galactic disks
exposed to UV radiation find that a gravitationally unstable cold
phase begins to appear above a fixed physical density threshold,
although the precise value is somewhat uncertain \citep{schaye04}.  If
we decreased the threshold, star formation would be more widespread
and some of the emission we attribute to self-shielded gas would
instead come from inside of galaxies (and be dominated by that due to
star formation).

For each star-forming particle, we convert the SFR reported by the
simulation to a Ly$\alpha$ luminosity via $L_\alpha = 10^{42} \, ({\rm
SFR}/{\rm M}_\odot {\rm yr}^{-1}) \ergs$, which is accurate to within
a factor of a few according to the stellar population synthesis models
of \citet{leitherer} for SF episodes with a Salpeter IMF, a stellar
mass range of $1$--$100 \msun$, and metallicities between $0.05 \Zsun
< Z < 2 \Zsun$ \citep{leitherer}, assuming that $\sim 2/3$ of ionizing
photons are converted to Ly$\alpha$ photons \citep{osterbrock89}.  Our
assumption of case-B recombination should be valid provided that most
ionizing photons are absorbed in the dense interstellar medium of the
galaxy.  The actual Ly$\alpha$ luminosity of a galaxy depends strongly
on the distribution of ionizing sources, the escape fraction of
ionizing photons, the presence of dust, and the kinematic structure of
the gas (e.g., \citealt{kunth03}), so this should be taken as no more
than a representative estimate.  Dust is particularly important,
because \lya photons must travel a long distance before scattering out
of resonance.  For example, only $\sim 20$--$25\%$ of bright
star-forming galaxies at $z=3$ have high-equivalent width \lya lines
(S00; \citealt{shapley03}), with the rest showing weak or no emission.
Because our simulations do not include this possibility, we expect our
results to overpredict the abundance of bright \lya emission by a
comparable factor.  Note that we assign the entire \lya luminosity to
the star-forming particle, which amounts to assuming that all ionizing
photons are absorbed close to their source.  In reality, some fraction
will escape to distant regions of the galaxy or even into the IGM.
Such detailed radiative transfer is beyond the capabilities of our
simulations, so we will simply note that \lya emission from star
formation could be more widely distributed.  In this case, although
the energy source is inside the galaxy, the \lya photons would still
allow us to probe the IGM.

In principle, some fraction of the interstellar medium of galaxies
could also produce diffuse \lya emission, especially if it is heated to
$T \sim 10^{4.2} \kel$.  Again, there is no way to follow the
detailed structure of the ISM in our relatively coarse cosmological
simulations.  We will therefore neglect such emission processes.

\subsection{Pumping from Higher Order Lyman Lines}
\label{pump}

The thick curves in Figure~\ref{fig:emiss} neglect \lya emission from
radiative pumping (i.e., absorption of a Lyman photon followed by a
radiative cascade that produces a \lya photon).  In Cloudy, we achieve
this using the ``no induced processes'' option.  One reason for this
choice is that we wish to exclude \lya photons emitted immediately
after absorbing a \lya photon from the background radiation field:
this process does not contribute to the \emph{net} luminosity of the
gas.  However, we should include absorption of higher Lyman line
photons that cascade to Ly$\alpha$.

Absorption of a Ly$\beta$ photon ($1s \rightarrow 3p$) cannot result
in emission of a \lya photon ($2p \rightarrow 1s$): the Ly$\beta$
photon puts the atom into the 3$p$ state, from which it must decay to
the 1$s$ or 2$s$ state.  Unfortunately, Cloudy assumes that all levels
with $n \ge 3$ have their orbital angular momentum states completely
mixed, which is only a good assumption for densities $n_{\rm H} \gg
10^8~{\rm cm}^{-3}$ \citep{pengelly64}, so it \emph{does} allow
unphysical conversion of Ly$\beta$ to Ly$\alpha$. This is another
reason why we choose the ``no induced processes'' option.

The selection rules do allow all higher Lyman lines to cascade
to Ly$\alpha$.  As an example we now estimate the importance of
Ly$\gamma$ pumping, in which the atom begins in the $4p$ state.  The
selection rules allow: (1) emission of another Ly$\gamma$ photon
through direct decay to the ground state, (2) emission of two photons
following decay to $2s$, or (3) the chains $4p \rightarrow 3s
\rightarrow 2p \rightarrow 1s$ and $4p \rightarrow 3d \rightarrow 2p
\rightarrow 1s$, which both produce \lya photons.\footnote{Note
however that not all $4p$ states allow decay to $3d$ because the total
angular momentum can only change by $\hbar$.}  Thus the fraction of
absorptions resulting in a \lya photon is
\bq
f_{\gamma,\alpha} = \frac{A_{4p,3d} + A_{4p,3s}}{A_{4p,3d} + A_{4p,3s}
+ A_{4p,2s} + A_{4p,1s}},
\label{eq:fgamma}
\eq
where $A_{x,y}$ is the Einstein $A$-coefficient between states $x$ and
$y$.  The result is $f_{\gamma,\alpha}=0.0444$, where we have averaged
over the $4p$ states.  This is the fraction for a single absorption;
however, if the medium is optically thick to Ly$\gamma$ photons, a
re-emitted Ly$\gamma$ photon will be repeatedly absorbed until it is
transformed into a \lya photon or a pair of $2s$ continuum photons.
To account for this possibility, we could set $A_{4p,1s}
\rightarrow 0$ in equation (\ref{eq:fgamma}), which yields
$f_{\gamma,\alpha}=0.2781$. 

The thin lines in Figure~\ref{fig:emiss} show the extra \lya
emissivity from Ly$\gamma$ absorption for several different densities
at $z=3$, assuming an optically thin medium.  We see that in this case
pumping is only a small correction for all densities; on the other
hand, if the medium were optically thick to Lyman line photons, the
correction could be comparable to the emissivity from recombinations
in the low-density IGM.\footnote{To accurately compute the pumping
contribution for a gas cloud that is optically thick in the Lyman
lines we would need to do a full radiative transfer calculation
because the Lyman line intensity would vary within the cloud.}
However, it is always much less than the emission from collisional
excitation in dense gas with $T \ga 10^{4.2} \kel$, which, as we will
see, dominates the total.  For simplicity, we will thus neglect this
process in the following and note only that it could increase the
luminosity of the low-density and/or low-temperature IGM by a small
factor.  Including even higher transitions than Ly$\gamma$ does not
change this general conclusion.  However, it should be noted that
pumping could be much more important in places where the intensity in
the Lyman lines significantly exceeds that of the mean background
radiation.

Finally, we note that the pumping contribution increases with
redshift, because in photoionization equilibrium the mean neutral
fraction increases with the proper density.

\section{IGM Characteristics}
\label{phase}

In this section we examine some general properties of \lya emission in
the simulations, including the phase diagram of the IGM (\S
\ref{phasediagram}), the underlying energy source for the \lya
radiation (\S \ref{energysource}), and the effects of galactic winds
(\S \ref{wind-phase}).  

\begin{figure*}[t]
\plotone{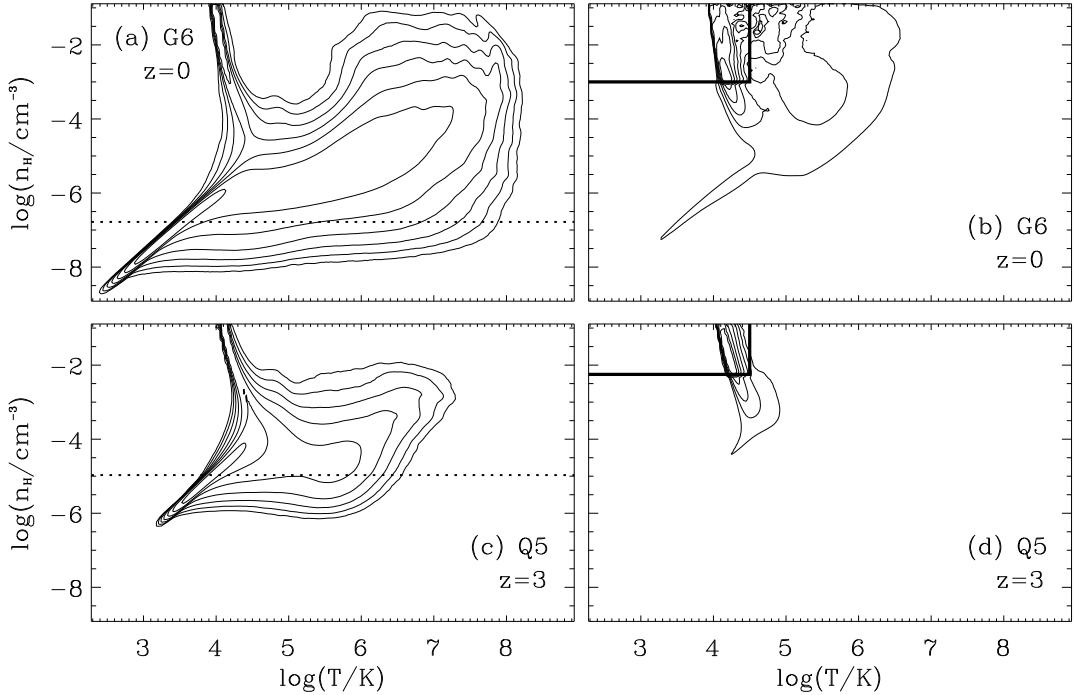}
\caption{Phase diagrams of the IGM.  \emph{(a):} Particle number
  density in the G6 simulation at $z=0$.  The dotted line shows the
  mean density at this epoch.  \emph{(b):} Particle number density
  weighted by Ly$\alpha$ emission.  The solid lines mark the
  self-shielded region of phase space; we assume CIE for gas in this
  region. \emph{(c), (d):} Same as the the top row, except for the Q5
  simulation at $z=3$.  All contours are equally spaced
  (logarithmically) in density, with an interval of 0.5 dex. }
\label{fig:phase}
\end{figure*} 

\subsection{Phase Diagram of the IGM}
\label{phasediagram}

Figure~\ref{fig:phase} shows $\nh$--$T$ phase diagrams for two
simulation outputs: G6 at $z=0$ (top row) and Q5 at $z=3$ (bottom
row).  The left column shows the particle number density (unweighted),
while the right column weights by the \lya luminosity of the
particles, assuming CIE for self-shielded gas.  F03 describe most of
the main features in detail.  The majority of particles lie along a
line connecting cool, underdense gas and moderately overdense, warm
gas.  This is the diffuse, photoionized IGM responsible for the \lya
forest \citep{katz96,hui97,schaye99,mcdonald01}.  Another set of
particles occupy a broad range of temperature at moderate to large
overdensities.  This gas is the shocked IGM (now known as the
``warm-hot IGM'' at $z \sim 0$).  At $z=3$, much of this gas has been
shocked by galactic winds, but by $z=0$ the nonlinear mass scale has
risen significantly and large-scale structure shocks provide most of
the heating \citep{cen99,dave01-whim,croft02, keshet03,furl04-sh}. A
third set of particles lies along a nearly vertical line at $T \sim
10^{4} \kel$, representing gas that has cooled and collapsed into
bound objects but is not currently forming stars.  We will refer to
this feature as the ``cooling locus.''  In simulations without winds,
the vast majority of these particles have not yet formed stars but
will do so in the relatively near future.  With winds, the
interpretation is more subtle (see \S \ref{wind-phase} below).  The
rapid decrease in the cooling rate at $T<10^{4.2} \kel$ forces this
gas to approach a constant asymptotic temperature near that value.
The high-temperature envelope of this locus appears because the
cooling time falls rapidly as the temperature increases.  Although
very few particles inhabit this region of phase space, their
emissivity is so large that they do appear in panel \emph{(b)}.

It is clear that most of the \lya emission comes from gas on the
cooling locus, especially at $z=3$, because this curve lies near the
peak of the \lya emissivity in CIE (note that the majority of this
region is in the self-shielded regime).  Because $\epsa$ is so
sensitive to temperature at $T \sim 10^{4.2} \kel$, the emission
characteristics will depend on the precise temperature along this
curve.  This is particularly evident in Figure
\ref{fig:phase}\emph{b}: the densest gas has $T<10^{4.1} \kel$, below
the peak of \lya emission, and so contributes only a small fraction of
the total luminosity.  In contrast, at $z=3$ the temperature is higher
and all of the gas contributes significantly.  As emphasized in \S
\ref{ss}, much of this cool gas will actually be self-shielded from
the mean metagalactic ionizing background (which is responsible for
most of the heating in the simulation) and hence may cool to a lower
temperature than the simulation allows; in this case the emissivity
could change dramatically.  The sensitive dependence of the emission
on the temperature is a direct consequence of assuming CIE.  If all of
the gas remains optically thin, $\epsa$ is relatively flat for $T \la
10^{4.5} \kel$ (see Fig.~\ref{fig:emiss}), so the densest gas at $z=0$
would still emit strongly. 

We do not show star-forming particles in Figure~\ref{fig:phase}.
Particles in the simulation form stars when $\nh > 0.129 \cmden$, so
they would all lie above the upper edge of these plots.

\subsection{The Energy Source of \lya Emission}
\label{energysource}

\citet{haiman00-lya} and \citet{fardal01} have suggested that
gravitational shock-heating can yield large \lya luminosities for
halos in the midst of collapse.  \citet{haiman00-lya} argued that the
gas would be shocked to the virial temperature of the halo and cool
rapidly to $T \sim 10^4 \kel$ through a combination of free-free and
line emission (but primarily higher excitation lines than hydrogen
Ly$\alpha$).  The accreted baryons would then be left out of
hydrostatic equilibrium and collapse to the center of the halo.
During this stage, the gas remains at $T \sim 10^4 \kel$ and radiates
its gravitational energy in \lya photons.  \citet{fardal01} found that
most of the gas was not shocked to high temperatures in their
simulation, so that a potentially even larger fraction of the
gravitational energy could be radiated in \lya photons.
\citet{birnboim03} reach similar conclusions from analytic arguments
and one-dimensional simulations, and \citet{keres04} confirmed the
conclusion in a detailed analysis of gas accretion in
three-dimensional cosmological simulations.  In either case, we expect
a large fraction of the emission to come from gas that has recently
been accreted, and naively that gravitational processes provide the
energy reservoir for the \lya emission.

\begin{figure*}[t]
\plotone{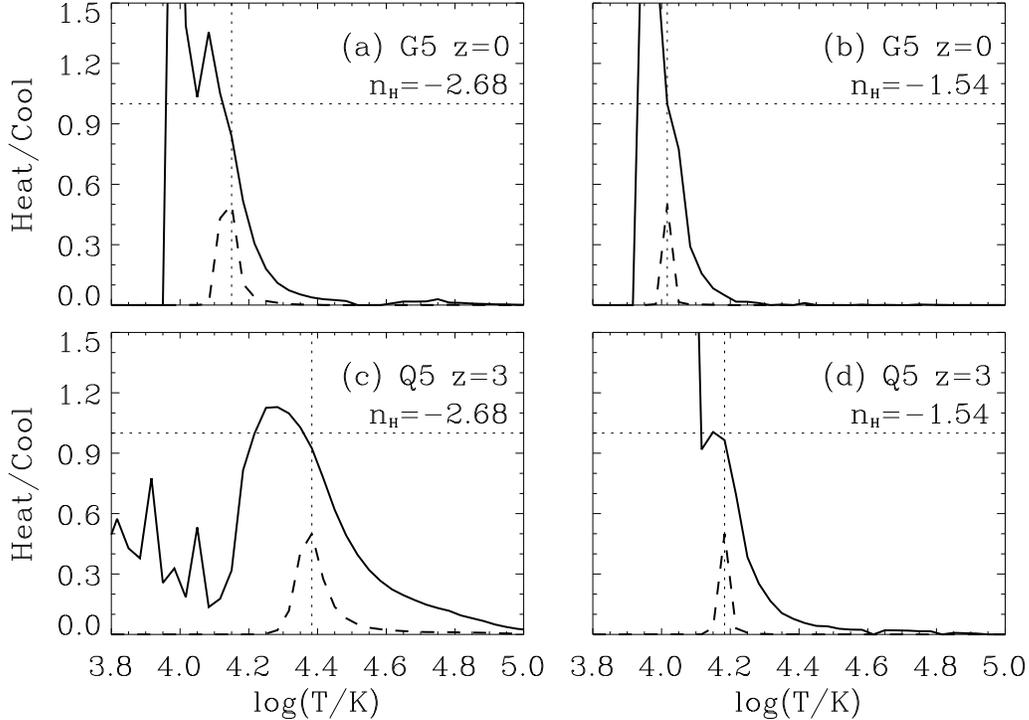}
\caption{Ratio of photoheating to total cooling along fixed density
  slices (solid curves).  The dashed curves show the particle
  distribution function at this density (with arbitrary
  normalization).  The vertical dotted line marks the point where the
  distribution function peaks; the horizontal line shows where heating
  and cooling just balance.}
\label{fig:energy}
\end{figure*} 

We can test this hypothesis with our simulations.  The primary
additional ingredient we have added is the ionizing background.  For
each particle in the simulation, we compute the photoheating rate as
well as the \emph{total} cooling rate, including radiative cooling and
adiabatic expansion.  The ratio of these two quantities yields the
fraction of \lya energy that can be directly attributed to the
ionizing background; the remaining energy can come from gravitational
processes (or any other, such as galactic feedback).  We show the
results in Figure~\ref{fig:energy}.  The upper solid curves show the
ratio of (radiative heating/total cooling) along fixed density slices.
The dashed curves show the particle distribution at that density (with
arbitrary normalization).  We have selected densities on the cooling
locus; note that $n_{\rm H,ss}=10^{-2.25},\,10^{-3} \cmden$ at
$z=3,\,0$.  The (heat/cool) ratio behaves erratically at low
temperature because there are few particles in this regime, many
of which have unusual histories (such as recent ejection by winds).
The distribution peaks where photoheating nearly cancels cooling.  The
particles are still cooling (albeit much more slowly because of the
ionizing background), especially in the low-density slice, and heading
toward the star formation threshold.  Clearly the particle
distribution is determined by the balance of photoheating and cooling.
Note that this is true even for self-shielded gas; as we have argued
in \S \ref{ss}, such behavior is an important limitation of the
simulations, as it represents an unphysical heat supply for such gas.

\begin{figure}[t]
\plotone{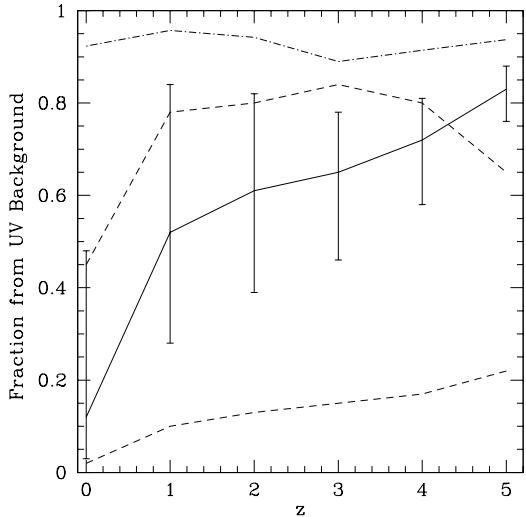}
\caption{Fraction of \lya radiation ultimately due to the ionizing
  background.  The solid curve shows the fraction for gas below our
  fiducial $n_{\rm ss}$; error bars show the fraction if we increase
  or decrease the threshold by 0.67 dex.  The dot-dashed curve assumes
  that all gas is optically thin.  The upper and lower dashed curves
  increase and decrease the ionizing background by one order of
  magnitude from our fiducial choice. }
\label{fig:ensource}
\end{figure}

The solid line in Figure~\ref{fig:ensource} shows the fraction of \lya
emission due to photoheating when we integrate over the entire
particle distribution (assuming $\epsa=0$ for self-shielded gas).  At
$z \ga 1$, photoheating provides somewhat more power than gravity or
winds (accounting for $\sim 2/3$ of the total emission).  On the other
hand, at $z \sim 0$, photoheating is negligible because the ionizing
background has fallen substantially.  These results imply that the
photoionizing background is important for optically thin gas; however,
with our fiducial $n_{\rm ss}$, such gas accounts for only $\sim
1$--$10\%$ of the total emission.  Whether photoionizations are
important for the total emission depends on our assumptions about cool
dense gas.

Unfortunately, it is difficult to self-consistently vary the ionizing
background, because it is built into the simulations.  In particular,
the equilibrium temperatures of cool dense gas particles depend on the
background.  A weaker background decreases the temperature and hence
the emissivity; a stronger background will usually increase the
emissivity unless it pushes the temperature over the peak.  To include
these effects rigorously would require radiative transfer to be part
of the simulation, which is not yet feasible.  We will therefore take
an approximate approach to varying the background.

First, we can vary our criterion for ``self-shielded'' gas.  This
tells us how important the ionizing background is as a function of
density for gas on the cooling locus.  It also mimics a situation in
which the local ionizing radiation field is larger than the mean.  The
error bars in Figure~\ref{fig:ensource} show what happens if we
increase or decrease $n_{\rm ss}$ by 0.67 dex.  Because the cool dense
gas has its temperature fixed almost entirely by the ionizing
background, increasing the threshold density also increases the
fraction of the emission from this mechanism.  The dot-dashed line
shows the fraction if $n_{\rm ss}=0.129 \cmden$, the SF threshold; in
this case, $\ga 90\%$ of the \lya emission comes from the ionizing
background.  We conclude that photoionization will strongly dominate
gravitational heating as the source of \lya emission, regardless of
redshift, if the local ionizing field is strong.

Second, we can change the amplitude of the ionizing background when
calculating the emissivity and heat/cool ratio (but \emph{not} the gas
temperature, because that is built into the simulation).  The dashed
curves show what happens if we increase or decrease $\Gamma$ by one
order of magnitude.  This choice also modifies $n_{\rm ss}$ by 0.67
dex (see equation [\ref{eq:sscol}]) as well as modifying the
distribution of emissivity by decreasing the importance of
recombinations following ionizations.  In particular, while
strengthening the ionizing background appears to have little effect
other than on $n_{\rm ss}$ (except at $z=5$), weakening it reduces the
importance of the UV background by a much larger amount.  This is
because with such a background most of the emission arises from hot,
dense gas at $T>10^{4.5} \kel$ for which collisional excitation
dominates completely (see Fig.~\ref{fig:emiss}).

With these results in mind, we now consider four cases for emission
from cool, dense gas.  First, it could remain optically thin because
of photoionization by embedded sources.  Then the simulation is
adequate: the temperatures are incorrect because the amplitude and
shape of the ionizing field change, but the temperature is relatively
unimportant for optically thin gas.  In this case, the dot-dashed
curve in Figure~\ref{fig:ensource} shows that photoionization
completely dominates.  Second, self-shielded gas could be invisible,
either because it cools rapidly or because of dust.  In this case the
simulation is exact (because only optically thin gas is observable),
and the photoionizing background dominates at $z \ge 1$ although not
to the exclusion of gravitational processes.  Third, self-shielded gas
remains visible and is heated by some unspecified source (this
corresponds to our CIE curves).  The simulation temperatures are
certainly unreliable in this case.  While optically thin gas would
still be powered by the ionizing background, the self-shielded gas
(and hence most of the emission) would by definition be powered by
some other mechanism.  Fourth, the gas could cool once it condenses
beyond $n_{\rm ss}$ and remain visible only during the initial
cooling.  Here, we have found that the ionizing field plays an
important (though not necessarily dominant) role in the initial
temperature (except at $z=0$).  Determining which of these
possibilities corresponds to reality requires more advanced
simulations that include self-consistent radiative transfer.

This does not imply that shocks are irrelevant to structure formation:
as \citet{haiman00-lya} point out, a large fraction of the shock
energy could be radiated through other transitions.  We briefly
consider the importance of shocks here.  We have randomly selected
luminous particles from the $z=2.75$ snapshot of the O3 simulation.
(As we will see in \S \ref{wind-phase}, winds complicate the
interpretation of particles on the cooling locus, so we have here
chosen our highest resolution simulation without them.)  For this
simulation, we have snapshots spaced at $\Delta z=0.05$ for
$z=2.75$--$3.25$ and at $\Delta z=0.5$ for $3.5 \le z \le 7$.  For
each particle, we find the largest temperature $T_{\rm max}$ and
entropy $K_{\rm max}$ (where $K= T/n^{2/3}$) it has experienced in
this set of snapshots.  (We exclude any snapshot at which the particle
formed stars, but these are extremely rare anyway.)  Note that this is
actually only a lower limit to the true maximum, because the outputs
are spaced fairly coarsely and the cooling time is short for large
temperatures and densities.

\begin{figure*}[t]
\plotone{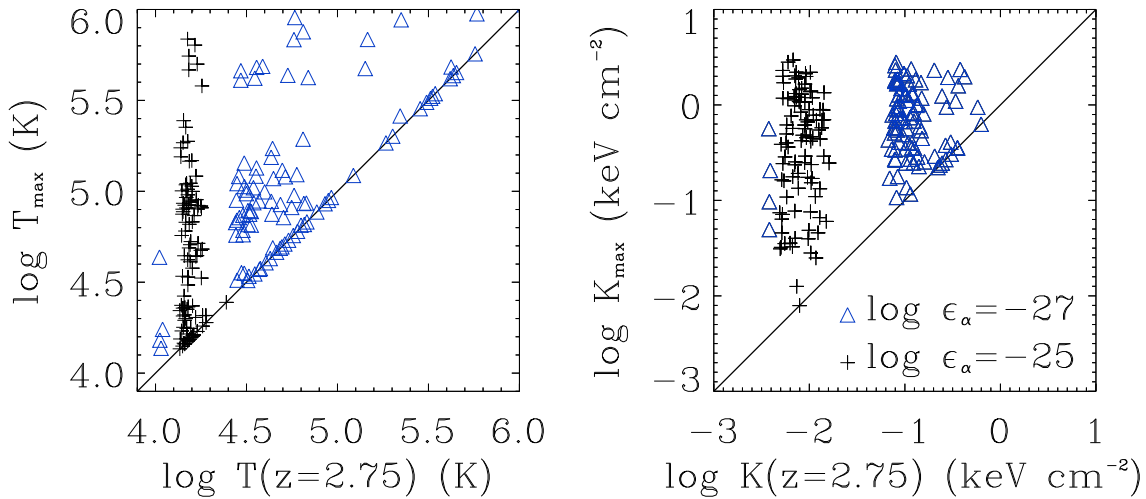}
\caption{\emph{Left:} Compares the current temperature of simulation
  particles to the maximum temperature achieved in the previous
  simulation outputs. Since we only have a finite number of outputs,
  the plotted maximum temperatures are lower limits to the true
  maxima. We have randomly selected 100 particles from the O3
  simulation at $z=2.75$ for each emissivity bin; the triangles and
  plus signs have $-27.25 < \log \epsa < -26.75$, and $-25.25 < \log
  \epsa < -24.75$, respectively (in units of$\emiss$).  The latter bin
  is composed of self-shielded particles (for which we assume CIE).
  \emph{Right:} Same, but for the entropy. }
\label{fig:trace}
\end{figure*}

Figure~\ref{fig:trace} compares the maximum and current $T$ and $K$
for particles with $-27.25 < \log \epsa < -26.75$ (triangles) and
$-25.25 < \log \epsa < -24.75$ (crosses, in units of$\emiss$).  The
former set has a relatively low emissivity and is not self-shielded;
the latter is well within the self-shielded regime.  We find that most
of the lower emissivity particles have $T_{\rm max} > T(z=2.75)$ (some
much greater than this level).  We expect virial temperatures of
$T_{\rm vir} \sim 10^5 (M/10^{10} \msun)^{2/3} \kel$ at $z=3$
\citep{barkana01}; most of the resolved star-forming halos in this
simulation have masses above $10^{10} \msun$.  Interestingly, most of
the particles remain at or below $2.5 \times 10^5 \kel$, the threshold
defined by \citet{keres04} to divide accreting gas into a ``cold'' and
``hot'' mode.  They argue that virial shocks are responsible for the
latter phase, while cold accretion occurs along filaments and is
subject only to weak shocks.  Our conclusions are consistent with
theirs; however, many of our halos have virial temperatures comparable
to the threshold, so the division is not as clean.  Like
\citet{keres04}, we do see some evidence for ``bimodality'' in the
temperature distribution, with very few particles having $T_{\rm max}
\sim 10^{5.5} \kel$.  This is likely because the cooling function has
a peak at these temperatures.  Typically, particles in this emissivity
range also have $K_{\rm max}$ several times larger than $K(z=2.75)$,
indicating a significant amount of radiative cooling regardless of the
accretion phase.

The higher luminosity particles show smaller maximum temperatures:
most have $T_{\rm max} \sim 10^{4.5}$--$10^5 \kel$, but many have
remained even cooler.  They have also lost more entropy, from initial
levels comparable to the other emissivity bin.  This is indicative of
substantial radiative losses.  One possible explanation for the
difference is that many of these particles were heated at $z>3.25$, so
that we are more likely to have missed the shock in our snapshots.
Alternatively, these may have been accreted through the cold mode
identified by \citet{keres04}.  Those authors argue that the cold mode
is ultimately responsible for most of the star formation at $z \sim
3$, and our results appear consistent with this conclusion.  We
emphasize again that, because of the short cooling times and limited
number of outputs, we actually only have lower limits on $T_{\rm max}$
and $K_{\rm max}$.

In summary, we cannot neglect the ionizing background (and any
additional contribution from local sources) when computing the \lya
emission from the IGM.  Unfortunately, its precise role depends on
self-shielding and local radiation sources, so a self-consistent
treatment will require radiative transfer in full cosmological
simulations, including its dynamical effects.  \citet{haiman00-lya}
and \citet{fardal01} neglected the ionizing background and argued that
the gas would contract after being shocked and radiate its
gravitational energy.  Instead, we find that the gas settles into a
quasistatic phase where photoheating nearly balances radiative
cooling.  This heat exchange becomes unphysical if (at some point
during the contraction) the gas becomes self-shielded (see \S
\ref{ss}).  The density at which this transition occurs will determine
how much energy the metagalactic ionizing background supplies compared
to gravitational collapse (and winds).

\subsection{Winds}
\label{wind-phase}

Most of the simulations that we study include a galactic wind
mechanism (see \citealt{springel-sf,furl04-metals} for detailed
descriptions of the wind implementation).  We might expect this
prescription to have important consequences for the \lya emission.  We
have therefore also examined some lower resolution runs with weak or
no winds (P3 and O3, respectively).  In this section we review the
effects of the wind prescription on the simulation particles.

The most important consequence of winds is to sharply reduce the star
formation rate: the SFR density is $(1.18,0.855,0.244) h^3
M_\odot\,{\rm yr}^{-1}\,\Mpc^{-3}$ at $z=3$ in the (O3, P3, Q3)
simulations, bringing the Q3 simulation into reasonable agreement with
observations \citep{springel03,hern03}.  The winds accomplish this in
two ways.  First, in the absence of ongoing accretion, they simply
deplete the gas in the star-forming phase by ejecting it into the IGM.
Second, winds interact with gas currently accreting onto halos.  In
small galaxies, a wind can severely reduce the accretion rates by
entraining gas and transporting it out of the halo or by shedding its
energy to gas in the outskirts of the halo.  In large galaxies, on the
other hand, winds remain bound to the halo and only weakly affect the
accretion rate. The halo gas rids itself of any wind energy through
cooling. In this case even the wind material itself can re-enter the
star forming phase after a short delay.

Thus, winds have a substantial effect on gas lying along the cooling
locus.  In their absence, nearly all of this gas is pristine (i.e.,
has not formed stars in the past).  If we trace the evolution of
individual particles, they slowly cool as their densities increase
until they are able to form stars.  If we turn on weak winds (i.e.,
those that are unable to escape the gravitational potential wells of
typical galaxies), we would expect more gas to lie on the cooling
locus because winds eject star-forming particles into the halo and
provide extra heat to the accreting gas.  Moreover, a significant
fraction of the dense gas would have formed stars in the past and thus
be metal-enriched.  As the winds strengthen, the amount of cool gas
should decrease again once they can completely eject particles from
galaxy halos.  However, some fraction of the wind particles will still
be trapped along ``accretion channels'' and fall back on the galaxies.

\begin{figure*}[t]
\plotone{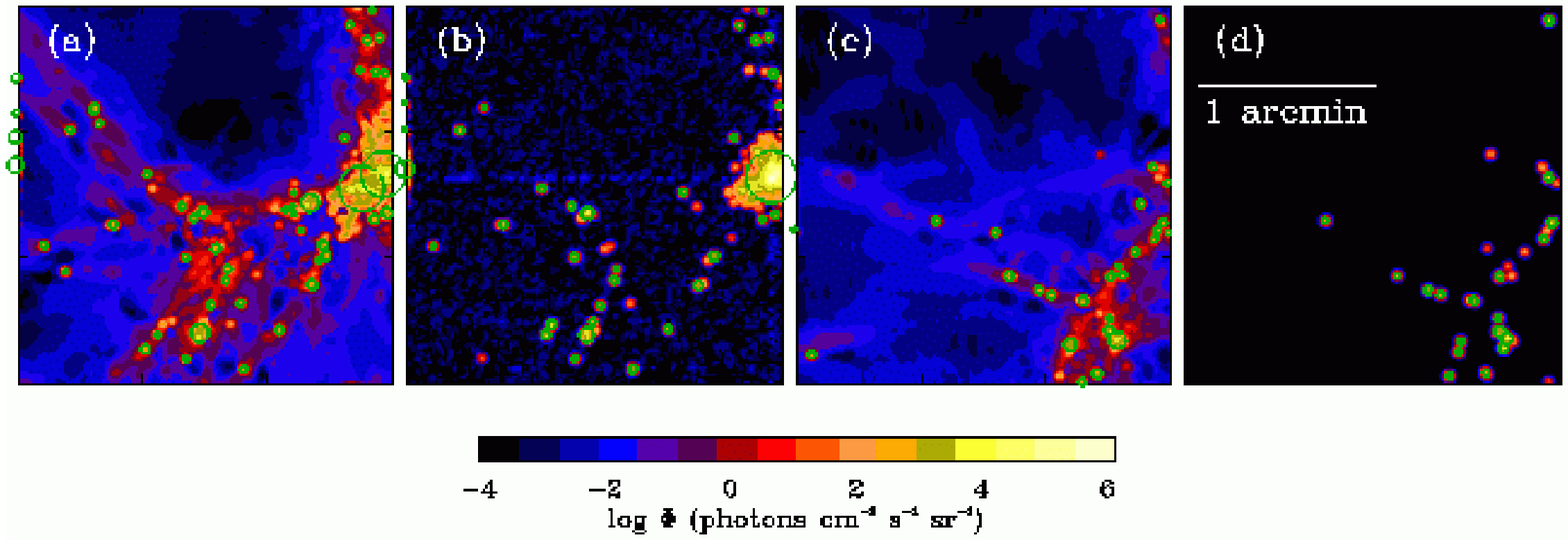}
\caption{Maps of Ly$\alpha$ emission from the IGM (\emph{a} and
  \emph{c}), assuming self-shielded gas is in CIE, and from star
  formation (\emph{b} and \emph{d}).  All are from the Q5 simulation
  at $z=3$.  Sources identified by SExtractor (with the threshold at
  $\Phi=100 \photunits$ or $9.6 \times 10^{-21} \sberg$) are outlined
  in green, with the size proportional to the source's area on the
  sky.  The slice shown in panels \emph{(a)} and \emph{(b)} is
  unusually rich; the other is more typical.  The field is $2.75
  h^{-1} \Mpc$ wide and $0.67 h^{-1} \Mpc$ thick (comoving; this
  corresponds to 1.2 \AA).  The angular resolution is $1\arcsec$.}
\label{fig:sez3map}
\end{figure*} 

We find precisely this trend in the simulations: the O3 simulation
actually has the least amount of gas along the cooling locus, even
though it has the highest SFR.  The Q3 simulation has slightly more
gas in this phase, while the P3 simulation has the most (apparently
these weak winds get trapped near the galaxies).  In both the Q3 and
P3 simulations, a significant fraction of high density, low
temperature particles have formed stars in the relatively recent past.
Thus winds have complex, and not always intuitive, implications for
\lya emission.  However, the effect on the fraction of gas lying along
the cooling locus is modest, varying by only a small amount.  The
upper left part of Figure~\ref{fig:phase}\emph{c} appears essentially
unchanged between the different wind prescriptions.

A more obvious effect is on the shocked, diffuse IGM at high
redshifts: without winds, this phase would be considerably weaker in
Figure~\ref{fig:phase}\emph{c}.  Of course, current cosmological
simulations cannot fully resolve the internal structure of galactic
winds.  In the nearby universe, winds are observed to have complex
multiphase media, including a hot diffuse interior, dense warm clouds,
and (in some cases) a cool dense shell of entrained matter.  These
components are largely unresolved in our simulations and could
strongly increase the \lya luminosity of the wind region and render
its geometry more accessible to observations.  The true effect of
winds may thus be more complex than simply adding gas to the hot,
low-density phase.

Winds are clearly complex phenomena, and their effects on \lya
emission are difficult to predict a priori.  It is obvious from
Figure~\ref{fig:phase}\emph{c} that the additional hot gas will make
a negligible difference to the emission.  We will examine the effects
on bright emission from the cooling locus in \S \ref{winds} below.

\section{Ly$\alpha$ Maps}
\label{maps}

We will now describe the characteristics of narrowband \lya maps in
our simulations.  In this section we will focus on qualitative
features and global statistics of the maps. We will discuss the
statistical properties of well-localized \lya emitters in \S
\ref{source}. Figure~\ref{fig:sez3map} shows two example slices of the
universe at $z=3$ (taken from the Q5 simulation).  The slice is 1.2
\AA\ thick (corresponding to $\Delta z=10^{-3}$ or $0.67 h^{-1} \Mpc$
comoving), and the field of view is $2.1\arcmin$ with angular
resolution $1\arcsec$ (these correspond to $2.75 h^{-1} \Mpc$ and
$21.6 h^{-1} \kpc$, both comoving).  Unless otherwise specified, all
of our $z=3$ maps assume these parameters.  Panels \emph{(a)} and
\emph{(c)} show the emission from IGM gas (assuming that the
self-shielded gas is in CIE), while \emph{(b)} and \emph{(d)} show
that from star-forming particles in the corresponding slices.  The
green circles outline distinct sources (see \S \ref{sizes} for
details).  The slice at right is typical for this redshift; the left
slice is among the richest few percent in the simulation.

These maps have a number of generic features.  First, bright \lya
emission from the IGM and from SF are highly correlated.
It is rare for one to appear without the other; as we have described
above, this is because most of the luminous IGM particles are either
cooling slowly to form stars or have recently felt the effects of
galactic winds.  The total luminosity of each region is usually
dominated by star-forming particles.  However, the IGM emission tends
to be distributed on somewhat larger scales than the SF.
Most of the emission from these coronae (especially in their cores)
comes from gas above the self-shielded threshold.  As we will show
below, these cores depend sensitively on our assumptions about
self-shielded gas.  Second, \lya emission is aligned along the cosmic
web of filaments, with low surface brightness emission (from optically
thin gas) connecting the coronae of bright emission surrounding
galaxies.  These properties remain true at all redshifts, though the
importance of the coronae varies strongly, as we shall show next.

\subsection{Redshift Evolution}
\label{zevol}

\begin{figure}[t]
\plotone{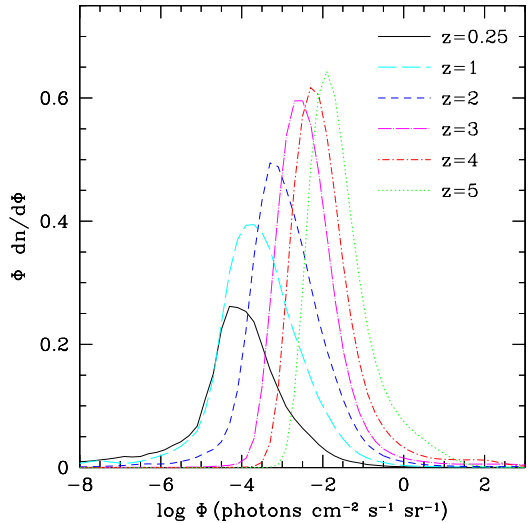}
\caption{Evolution of ${\rm d}n/{\rm d}\ln\Phi$.  The pixels are $21.6
  h^{-1} \kpc$ wide and $0.67 h^{-1} \Mpc$ thick (comoving); this is
  $1\arcsec$ wide and 1.2 \AA \ thick at $z=3$.  Note that the area
  under the $z=0.25$ curve is less than unity because $\sim 30\%$ of
  pixels do not include any simulation particles.}
\label{fig:zevol-full}
\end{figure}

Figure~\ref{fig:zevol-full} shows how the statistical properties of
the maps evolve with cosmic time.  For each redshift, we show the
probability distribution function ${\rm d}n/{\rm d}\ln\Phi$ of pixel
surface brightness $\Phi$ computed from 120 slices of the highest
resolution simulation available at that redshift.  In order to better
interpret the physical processes as they evolve, we have held the
comoving spatial resolution and slice thickness constant with
redshift; note then that the angular resolution and spectral width of
the observations differ.  (They equal the choices of Figure
\ref{fig:sez3map} at $z=3$.)  We have assumed CIE for the
self-shielded gas, although the choice makes no visible difference to
this Figure.  At all redshifts, ${\rm d}n/{\rm d}\ln\Phi$ peaks at a
relatively small surface brightness characteristic of the median
cosmic density.  In the highly-ionized low-density IGM, recombinations
dominate over collisional excitation (see Fig.~\ref{fig:emiss}), so
the mean surface brightness can be estimated from the recombination
rate \citep{hogan87,gould96}.  It is:
\begin{eqnarray}
\frac{{\rm d} \Phi}{{\rm d} \lambda_{\rm obs}} & = & \frac{\eta
  \dot{n}_{\rm rec}}{4 \pi 
  (1+z)^3} \, \frac{{\rm d} r_{\rm phys}}{{\rm d} z} \, \frac{{\rm d}
  z}{{\rm d} \lambda_{\rm
  obs}} 
\label{eq:lyaestimate} \\
\, & \approx & 3 \times 10^{-3} \left( \frac{\Omega_b
  h^2}{0.02} \right)^2 \frac{(1+z)^2}{h(z)} \nonumber \\ 
& & \times \delta^2 h^{-1}
  \photfluxA, 
\nonumber
\end{eqnarray}
where $\eta \approx 0.42$ is the fraction of recombinations that
produce a \lya photon, $\dot{n}_{\rm rec}$ is the recombination rate,
$h(z) = H(z)/H_0,$ and $\delta$ represents the effective overdensity
(i.e., the density relative to the cosmic mean) of the observed pixel.
Here we have used $\alpha=4.2 \times 10^{-13} \ {\rm cm}^{3} \ {\rm
s}^{-1}$ for the (case-A) recombination rate, appropriate for gas with
$T=10^4 \kel$.  (This estimate does not apply to the bright coronae,
which are not necessarily highly ionized and which have broken off
from the cosmic expansion.)  We see that $\Phi$ \emph{increases} with
redshift because the mean density and recombination rate increase
rapidly.  This equation yields a reasonable estimate of the peak of
${\rm d}n/{\rm d}\Phi$, although note that the appropriate $\delta$
changes as structure formation progresses.

\begin{figure}[t]
\plotone{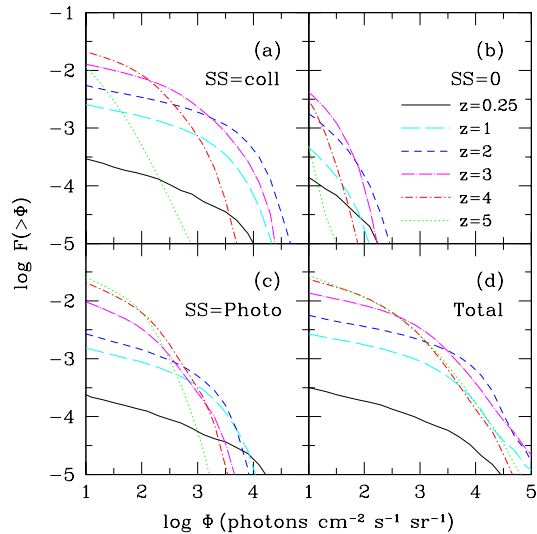}
\caption{Evolution of pixel brightness distribution. We show the
  fraction of pixels above a given surface brightness. The pixels are
  $21.6 h^{-1} \kpc$ wide and $0.67 h^{-1} \Mpc$ thick (comoving);
  this is $1\arcsec$ wide and 1.2 \AA \ thick.  \emph{(a):} IGM
  emission if self-shielded gas is in CIE.  \emph{(b):} IGM emission
  without self-shielded gas. \emph{(c):} IGM emission if all gas is
  optically thin to ionizing radiation.  \emph{(d):} Total Ly$\alpha$
  emission from both the IGM (including self-shielded gas as in panel
  \emph{a}) and SF.}
\label{fig:zevol}
\end{figure}

Unfortunately, the probability distributions of the pixel flux peak
many orders of magnitude below realistic detection thresholds,
regardless of redshift.  We therefore focus on the bright end of the
distribution in Figure~\ref{fig:zevol}, where we show the fraction of
pixels above a given threshold surface brightness, $F(>\Phi)$.  The
first three panels show emission from the IGM if self-shielded gas is
in CIE, if it has $\epsa=0$, and if it is optically thin to the
ionizing background.  The fourth shows the total emission from the map
(including SF, and assuming CIE for the self-shielded gas).

Looking first at panel \emph{(a)}, we see that the bright pixels
evolve in a way that differs qualitatively from the low-density IGM.
The brightest emission peaks at $z \sim 2$--$3$ and decreases rapidly
toward high and low redshifts.  The decline occurs when the cooling
locus shifts to low temperatures because the ionizing background is
weak and/or soft.  As a consequence, $\epsa$ declines rapidly if we
assume CIE, so the maximum pixel brightness falls.  The effect is
particularly severe at $z\sim5$, where the temperature of this
component is $T \approx 10^{4.0} \kel$ and self-shielded gas is
essentially invisible.

Comparison to panel \emph{(b)}, where we set $\epsa=0$ for
self-shielded gas, shows that nearly all of the emission with $\Phi
\ga 100 \photunits$ comes from self-shielded gas.  If this component
cannot emit (either because of strong dust absorption or because the
simulation overestimates its temperature), the \lya coronae become
orders of magnitude weaker.  Clearly the choice of $n_{\rm ss}$ is
crucial in the no emission case.  If our prescription is pessimistic
(because of embedded sources, for example), the coronae rapidly
strengthen.  F03 show an explicit example of how the $z=0$ statistics
change with the self-shielding cut.  Thus observations of diffuse \lya
emission around galaxies can strongly constrain the behavior of cool
gas in these environments.

Figure~\ref{fig:zevol}\emph{c}, on the other hand, shows the emission
if we assume that \emph{all} the gas is optically thin to ionizing
radiation.  Although \citet{schaye01-damp,schaye} has shown that
self-gravitating gas clouds become self-shielded from the
metagalactic background at these densities, most of these particles
surround star-forming galaxies.  As a result, the local ionizing field
could be much stronger than the mean, in which case the gas may remain
optically thin.  In this scenario, the maximum brightness decreases,
because optically thin gas is always highly ionized and the CIE
collisional excitation peak disappears.  However, the fraction of
moderately bright pixels remains nearly the same, because the same
particles emit (and remain reasonably bright).  The exception is at
$z=5$, where the optically thin case has much stronger emission. This
is because the temperature dependence of $\epsa$ is much weaker in the
optically thin case, so our predictions are much less sensitive to the
location of the cooling locus.

The final panel shows that emission due to SF dominates the brightest
end of ${\rm d}n/{\rm d}\Phi$ at all redshifts.  This is not
surprising, as SF produces \lya photons efficiently.  We find that
bright emission from SF rises to $z\sim3$ and then declines slowly at
higher redshifts.  Although the SFR slowly increases to $z \sim 6$ in
these simulations \citep{springel03,hern03}, the evolution flattens
beyond $z \sim 3$ because the increasing cosmic distance causes the
maximum surface brightness to decline.  On the other hand, note that
the fraction of moderately bright pixels does not change significantly
between panels \emph{(a)} and \emph{(d)} (except at $z=5$).  This is
because the IGM \lya emission is more extended (and fainter) than the
star-forming regions, so IGM emission dominates the sky coverage at
all but the highest surface brightnesses.  Of course, we have assigned
all of the \lya photons from SF to the host particles.  Some fraction
will actually escape, spreading this component over a wider area.
This mechanism will help to boost the IGM emission, even if the
ultimate energy source is (photoionization from) SF.  Moreover, we
have neglected dust absorption, which will likely remove a large
fraction ($\ga 75\%$ according to S00) of the SF \lya photons.  Thus
the bright tail may not be as pronounced as indicated in
Figure~\ref{fig:zevol}\emph{d} and IGM emission (which probably
suffers less dust extinction) may be even more important.

\begin{figure}[t]
\plotone{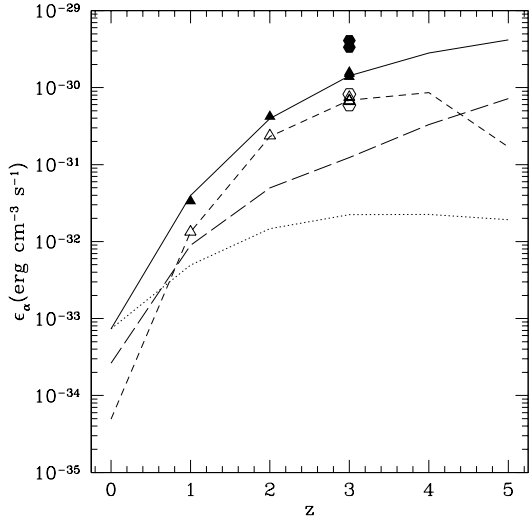}
\caption{The evolution of the volume-averaged emissivity
  $\epsilon_\alpha$ in physical units.  The solid line shows the total
  emissivity (assuming CIE for self-shielded gas). The short- and
  long-dashed curves show $\epsa$ from the IGM if we assume CIE for
  self-shielded gas and that all gas is optically thin, respectively.
  The dotted line shows the total emissivity per \emph{comoving}
  volume.  Triangles show the emissivity estimated from lower
  resolution simulations, while hexagons show it with different wind
  prescriptions.  Filled and open symbols refer to total and IGM
  emission, respectively.  }
\label{fig:bkgd}
\end{figure}

Figure~\ref{fig:bkgd} shows the redshift evolution in a different way.
Here we plot the volume-averaged emissivity in the \lya line.  The
solid line shows the total $\epsa$ (including both SF and the IGM,
assuming CIE for self-shielded gas).  The short- and long-dashed
curves show $\epsa$ from the IGM if we assume CIE for self-shielded
gas and that all gas is optically thin, respectively.  The dotted
curve again shows the total emissivity, but this time in comoving
units.  We have used the highest resolution simulation available at
each redshift to construct the curve; the triangles show estimates
from lower resolution simulations with the same parameters.  The
predictions have converged nicely.

One of the crucial lessons of Figure~\ref{fig:bkgd} is the difference
between the two dashed curves.  Evidently self-shielded regions
dominate the IGM emissivity.  The optically thin emissivity closely
traces the total (and hence the SFR).  Again, this is because $\epsa$
depends only weakly on temperature for optically thin gas, so the
coronal emission depends (to zeroth order) only on the rate at which
galaxies accrete gas (or eject gas through winds), which is itself
roughly proportional to the SFR.  In contrast, in CIE $\epsa$ depends
sensitively on temperature.  At $z\sim2$--$3$, quasars make the
ionizing background rather hard and intense, so the cooling locus
shifts to high temperature and the CIE emissivities are large.  But at
$z=0$ and $z=5$, the temperatures are small and CIE emissivities fall
below the optically thin values.  We will focus on $z=3$ in much of
the following, and we urge the reader to keep in mind that the CIE
assumption is the most optimistic at this epoch.

\subsection{Simulation Resolution}
\label{simres}

\begin{figure}[t]
\plotone{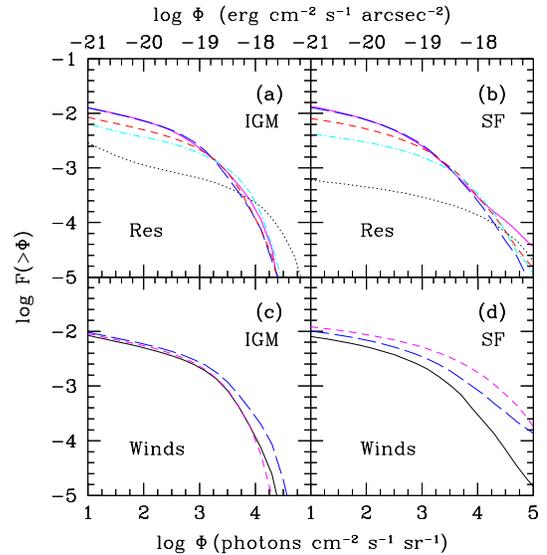}
\caption{Bright end of the pixel surface brightness distribution at
  $z=3$; the pixel scale is the same as in Figure~\ref{fig:sez3map}.
  \emph{(a):} Varying simulation resolution, with self-shielded gas in
  CIE.  The dotted, dot-dashed, short-dashed, long-dashed, and solid
  curves are for the G6, D5, Q3, Q4, and Q5 simulations, respectively.
  \emph{(b):} Same as \emph{(a)}, except for emission from star
  formation.  \emph{(c):} IGM emission if self-shielded gas is in CIE.
  The solid, long-dashed, and short-dashed curves have strong, weak,
  and no winds, respectively.  \emph{(d):} Same as \emph{(c)}, except
  for emission from SF.}
\label{fig:cdfmisc}
\end{figure}

We now consider the degree to which the finite simulation resolution
affects our results.  Figures \ref{fig:cdfmisc}\emph{a} and
\ref{fig:cdfmisc}\emph{b} show $F(>\Phi)$ for five different
simulations at $z=3$.  The two panels show results for IGM emission
(assuming self-shielded gas is in CIE; other cases yield similar
results) and that due to SF, respectively. All the parameters of the
slices are the same as in Figure~\ref{fig:sez3map}.  We also use the
same number of slices for each simulation.  Thus we only sample a
small fraction of the volume of the D5 and G6 boxes, and we avoid
contamination from rare objects that do not appear in the smaller
boxes.  The figure shows that our two highest resolution runs, Q4 and
Q5, appear to have converged.  However, the lower resolution runs
(especially G6) show an excess of extremely bright pixels together
with a deficit of moderately bright pixels, for both the IGM and star
formation.  This may seem surprising, because \citet{springel03}
showed that the global SFR has converged between all of these
simulations.  We find deeper convergence problems because we require
the spatial distribution of material within halos.  In the lower
resolution boxes, small halos are not as well-resolved and have SF
(and cool gas) confined to only a few particles that fill more compact
central regions.  Thus the lower resolution simulations can have
``cuspier'' surface brightness distributions that underestimate the
physical extents of the bright \lya regions.  The lower resolution
simulations also miss small halos, of course.  
On the other hand, it is possible that the extra large-scale power in
the D5 and G6 simulations genuinely increases the number of bright
pixels, by (for example) concentrating the star formation in smaller,
highly-biased regions.

\subsection{Winds}
\label{winds}

As described in \S \ref{wind-phase}, winds have a substantial impact
on the SF properties of the simulation, and we would naively expect
them to have comparable effects on the \lya
emission. Figure~\ref{fig:cdfmisc}\emph{d} shows that winds do indeed
have substantial implications for the \lya emission from star-forming
particles: the simulations with weak and no winds have nearly an order
of magnitude more pixels with $\Phi>10^4 \photunits$ than the
simulation with strong winds.  However, Figure
\ref{fig:cdfmisc}\emph{c} shows that the effects on the IGM emission
are quite modest.  (Here we have assumed CIE for the self-shielded
gas, but the differences are small for the other treatments of
self-shielded gas as well.)  The simulations with strong and no winds
have nearly the same $F(>\Phi)$, while weak winds strengthen the
emission by only a small amount.  This follows the trends expected
from the discussion in \S \ref{wind-phase}, in which we argued that
weak winds have the strongest effect on the cooling locus because the
winds cannot escape the potential wells of their host galaxies.  It is
not, however, obvious why the net effect should be so small.  It is
unclear whether there is a deeper reason for this behavior.  We do
note that the nature of the bright particles changes between the three
simulations.  Without winds, the vast majority of the emission comes
from particles that have not yet formed stars and contain no metals.
With strong winds, fewer than $50\%$ of bright particles are pristine,
because a number of formerly star-forming particles have mixed with
the gas in the cooling locus.  

We also note again that the simulations do not resolve the internal
structure of the winds.  If the wind medium fragments into dense
clumps inside of a more rarefied interclump medium, the total amount
of emission could change substantially.  The clumps would initially
have larger emissivity because of their increased recombination rate;
as a result, they would cool more rapidly and could eventually pass
below the \lya excitation threshold.  As with self-shielded gas, the
resulting emissivity depends sensitively on the equilibrium
temperature, which would in turn depend on the ionizing background,
continued heat input from the wind, and conduction from the hot
intercloud medium.  More detailed simulations of individual winds are
necessary to resolve these processes.

Figure~\ref{fig:bkgd} also compares the volume-averaged $\epsa$ across
wind simulations.  The hexagons show the mean emissivity for the P3
(weak wind) and O3 (no wind) simulations.  While the IGM emission is
approximately constant in the three cases, the mean SFR rises by a
factor of 4--5 from the strong wind case, increasing the total $\epsa$
by a large amount.

\subsection{Line Widths}
\label{vel}

We now examine the velocity widths of these \lya lines.  To estimate
the distribution, we construct simulation maps as before.  For each
particle, we also record its velocity (including both the Hubble flow
and the peculiar velocity) and velocity dispersion (assuming pure
thermal broadening).  From these we compute the flux-weighted velocity
dispersion $\sigma$ (i.e., the standard deviation of the velocity
distribution) of each pixel in the maps, using the same pixel sizes as
in Figure~\ref{fig:sez3map}.  Note that $\sigma$ includes thermal
broadening, peculiar velocities, and the Hubble flow between particles
(but \emph{not} the Hubble flow gradient within individual particles,
because that is negligible for bright, compact particles).

\begin{figure}[t]
\resizebox{9cm}{!}{\includegraphics{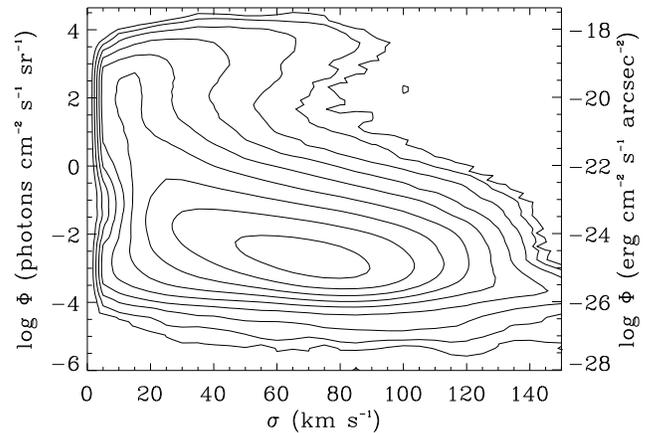}}
\caption{The distribution of pixel linewidths $\sigma$ (which we take
  to be the standard deviation of the line profile), including
  peculiar velocities and thermal broadening, in the Q5 simulation at
  $z=3$.  We do \emph{not} include radiative transfer effects.  The
  angular resolution is $1\arcsec$ ($21.6 h^{-1}$ comoving kpc) and
  the slices are 1.2 \AA \ thick ($0.67 h^{-1}$ comoving Mpc); $100
  \kms$ corresponds to $0.4$ \AA \ here. The contours are spaced 0.5
  dex apart.}
\label{fig:vel}
\end{figure}

Figure~\ref{fig:vel} shows the distribution of $\sigma$ per pixel
(including IGM emission only, and assuming CIE for the self-shielded
component) as a function of surface brightness $\Phi$.  Note that the
slice has a velocity thickness of $\approx 300 \kms$; uniform Hubble
flow would yield $\sigma \approx 80 \kms$.  This is indeed the peak of
the distribution in the low-density IGM.

We see that bright pixels are narrow. Essentially all pixels brighter
than 10 photons s$^{-1}$ cm$^{-2}$ sr$^{-1}$ have $\sigma \la 100
\kms$.  At modest brightness ($\sim 100 \photunits$), the distribution
peaks at $\sigma \approx 10$ to 15 \kms, which can be accounted for by
thermal broadening at the temperature of the cooling locus. As the
brightness increases, the distribution broadens to somewhat larger
velocity dispersions, indicating that peculiar motions become
relevant.  This is probably because the most luminous objects are
associated with more massive halos with large internal velocity
dispersions (see \S \ref{lumfcn}).  Note, however, that the simulation
has only limited resolution for the internal structure of halos, so
there may be extra broadening due to rotation, disc formation, etc.
Moreover, the coronae should be optically thick to \lya photons.  In
this case the photons can only escape by scattering into the wings of
the line.  Thus the velocities reported here will underestimate the
true linewidth.  For some geometries, radiative transfer can even
change the line's shape (e.g., \citealt{zheng02} and references
therein).  The best we can say is that the source regions are
intrinsically narrow before the effects of radiative transfer have
been included.

\vspace{0.3in}

\subsection{Comparison with the Diffuse Backgrounds}
\label{bkgd}

The observability of \lya emission, especially at the low surface
brightness levels relevant for most of the IGM, depends ultimately on
the emission strength relative to the diffuse background light.  This
background can be divided into four components: terrestrial airglow,
zodiacal light, diffuse galactic emission (i.e., scattered starlight),
and the extragalactic background.  The first is obviously unimportant
for space-based observations and we will not discuss it further here.
The other three are unavoidable.

The best measurements of the backgrounds level in the relevant regime
come from \citet{bernstein02}.  They find total backgrounds of $J_{21}
\sim 16$--$50$ over the range 3000-8000 \AA, where $J_\nu = J_{21}
\times 10^{-21} \junits$.  Of this total, $\ga 85\%$ comes from the
zodiacal light, which can be modeled as reflected sunlight.  They
estimate that diffuse galactic emission has $J_{21} \sim 0.5$ along
clear lines of sight through the Galaxy.  The extragalactic background
(from galaxies with $V > 23$ AB magnitudes) is then
$J_{21}=(2.0,1.3,1.1)$ at $(3000,5500,8000)$ \AA, with fractional
uncertainties of $\sim 50\%$ in each case.  These are a few times
smaller than the prediction of \citet{haardt01} at $z=0$.

Lines with fluxes above these levels would be straightforward to
detect in a background-limited observation.  In our units, the
integrated background over a velocity width $\Delta v$ is
\bq
\Phi_{\rm bkgd} = 10 \, J_{21} \left( \frac{\Delta v}{20 \kms} \right)
\photunits.
\label{eq:phibkgd}
\eq
As we have seen in \S \ref{vel}, bright and moderately bright pixels
typically have $\sigma \sim 10$--$50 \kms$, at least if we neglect
radiative transfer.  Thus we expect that features with $\Phi \ga 100
\photunits$ should be visible against the zodiacal light, and features
with $\Phi \ga 10 \photunits$ should be visible against the other
backgrounds.  This emphasizes the importance of high spectral
resolution observations, because the background increases as the
spectral resolution decreases.  The typical line widths are $\sim 0.16
(\Delta v/20 \kms) [(1+z)/4]$ \AA\ (again neglecting radiative
transfer effects).  We therefore recommend spectral resolution be a
priority in any observations designed to probe the lowest surface
brightness objects.

In fact, so long as the background light is either \emph{spectrally}
smooth or can be modeled to high precision, observations can in
principle extend far below the limits stated above.  For example,
spectral features in the zodiacal light mirror those in the solar
spectrum and can be modeled fairly well, as \citet{bernstein02} did.
The \lya emission would then appear as (weak) fluctuations on the
residual background.  We emphasize that the background need not be
spatially uniform, so long as the spectrum in each pixel is either
smooth or can be modeled.  \citet{zald04} discuss in some detail
conceptually similar ``foreground-cleaning'' techniques for
low-frequency radio observations (for applications, see also
\citealt{furl04a,furl04b}).

\section{Source Characteristics}
\label{source}

\subsection{Luminosity Functions}
\label{lumfcn}

We will now consider individual \lya sources.  We first divide the
simulation particles into gravitationally bound groups using a
friends-of-friends group finder with linking length equal to $20\%$ of
the mean interparticle separation and a minimum group size of 32
particles.  The linking is performed on the dark matter, with each
baryonic particle assigned to its nearest dark matter neighbor.  We
compute the \lya luminosity of each source as described in \S
\ref{emiss}.

\begin{figure}[t]
\resizebox{9cm}{!}{\includegraphics{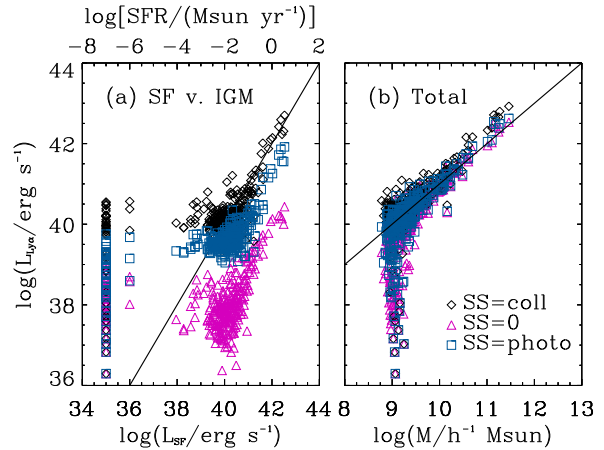}}
\caption{Characteristics of bound groups in the Q5 simulation at
  $z=3$.  \emph{(a):} Compares the IGM luminosity to that from stars
  in each group.  If the group has a vanishing current star formation
  rate (SFR) but contains star particles, we set $L_{\rm SFR}=10^{35}
  \ergs$.  If it has no star particles, we set $L_{\rm SFR}=10^{36}
  \ergs$.  The solid line shows $L_{\rm SFR}=L_{\rm Ly\alpha,\, IGM}$.
  \emph{(b):} Total Ly$\alpha$ luminosity (including SF) as a function
  of group mass, with the solid line showing the slope if $L_{\rm
  Ly\alpha} \propto M$.  The different sets of points correspond to
  our different assumptions about self-shielded gas.  In each case we
  have randomly selected 300 well-resolved groups.}
\label{fig:scatter}
\end{figure}

Figure~\ref{fig:scatter} shows some of the characteristics of groups
in the Q5 simulation at $z=3$.  Panel \emph{(a)} compares the \lya
luminosity from the IGM to that from SF.  The diamonds, triangles, and
squares assume self-shielded gas is in CIE, has $\epsa=0$, and remains
optically thin, respectively.  In each case we have randomly selected
300 well-resolved groups to display (i.e., they each have at least 10
times the minimum group mass).  The solid line shows $L_{\rm
Ly\alpha}=L_{\rm SFR}$.  For groups with no ongoing SF, we set $L_{\rm
SF}=10^{35} \ergs$ if the group has star particles and $L_{\rm
SF}=10^{36} \ergs$ otherwise.\footnote{The latter set is composed of
groups near the resolution limit.}  The scatter increases toward lower
luminosities, at least in part because accretion and star-formation
are more stochastic in halos that are not as well-resolved.  Clearly,
the SFR and the IGM luminosity are roughly proportional in all
treatments, as one would expect if the IGM emission comes from gas
that will soon become available for star formation (provided that the
gas temperature does not vary strongly between the different groups).
The wind prescription also forces the ejected mass to be proportional
to the SFR, so it leads to the same relation.  $L_{\rm SF}$ typically
exceeds the IGM luminosity except in the most optimistic case (which
is CIE at $z=3$).  The emission from optically thin gas is typically
only $\sim 0.01$--$0.05 L_{\rm SF}$; if self-shielded gas does not
emit, the luminosities of most sources would be completely dominated
by the emission from the associated SF.  Note, however, that the
symbols on the left indicate that some groups have sizable \lya
luminosities even though they are not currently forming stars.  Of the
groups above our mass threshold, $13\%$ have no ongoing SF but do have
star particles, while another $1.7\%$ have neither SF nor star
particles.

Figure~\ref{fig:scatter}\emph{b} shows the \emph{total} \lya
luminosity as a function of group mass.  This is nearly independent of
our treatment of self-shielded gas because SF dominates in most cases,
except of course for objects without ongoing SF (which are normally
near the mass threshold).  As we approach this threshold, the scatter
increases rapidly because SF is stochastic in poorly-resolved halos.
The solid line shows $L_{\rm Ly\alpha} \propto M$.  This appears to be
slightly too shallow to match the simulations, but only by a small
amount.  Figure~7 of \citet{springel03} shows that, in the mass range
we consider here, the SFR is also roughly proportional to mass, so
this is not surprising.  For higher mass halos (taken from the D5
simulation, for example), the relation steepens slightly.
Interestingly, however, the IGM emission \emph{declines} slightly
relative to that from SF in massive halos.  If we had included only
IGM emission, $L_{\rm Ly\alpha} \propto M$ remains a reasonable fit,
although the relation is slightly steeper if we exclude emission from
self-shielded gas.

\begin{figure*}[t]
\plotone{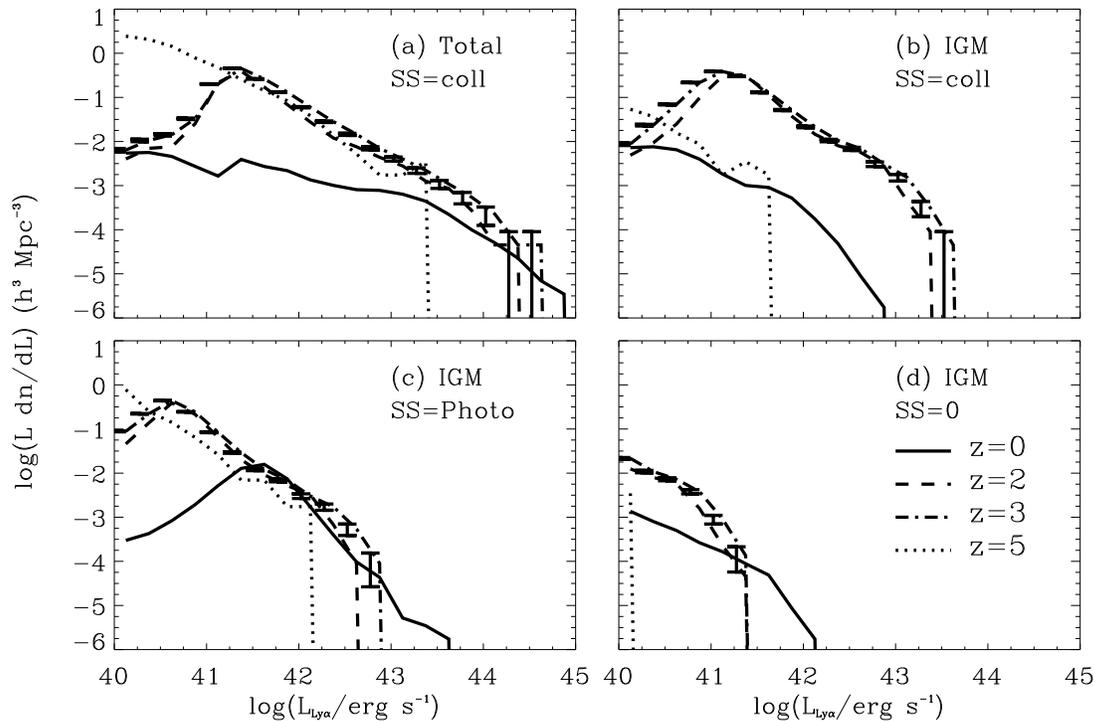}
\caption{The evolution of the luminosity function of groups.
  \emph{(a):} Total luminosity function, including both IGM (assuming
  CIE for self-shielded gas) and SF. \emph{(b):} IGM
  emission only, assuming CIE for self-shielded gas.  \emph{(c):} IGM
  emission only, assuming all gas is optically thin. \emph{(d):} IGM
  emission only, assuming $\epsa=0$ for self-shielded gas.  All distances
  are comoving.  Error bars (shown only for $z=3$) are Poissonian. }
\label{fig:lumfuncz}
\end{figure*} 

This contrasts with the predictions of \citet{haiman00-lya} and
\citet{fardal01}, who argue that cooling following gravitational
collapse should yield a total luminosity proportional to the binding
energy of the halo, or $L_{\rm IGM} \propto M^{5/3}$.  We find that
the photoionizing background actually provides most of the energy in
our calculations; previous work had neglected this component.  In this
case, the \lya luminosity should depend primarily on the mass of cool
gas in each group, so we naturally expect $L_{\rm Ly\alpha} \propto
M$.  The fraction of energy that can be traced to the ionizing
background is smallest if $\epsa=0$ for self-shielded gas, which
explains why the relation is steepest in that case.  Of course, some
of this energy exchange is unphysical, depending on if and when the
gas becomes self-shielded (see \S \ref{energysource}).  If most of the
contraction occurs while gas is shielded, the only available heat
source would be gravitational and we would expect $L_{\rm Ly\alpha}
\propto M^{5/3}$.  The difference could also be due to our wind
prescription.  Winds eject gas more efficiently in smaller halos and
provide a new source (beyond accretion) for the emitting particles.
However, although decreasing the strength of the winds increases
$L_{\rm SF}$ by a factor of a few, it has little effect on the IGM
emission.  The IGM luminosity of small groups increases slightly for
the weak wind case, but it does not appear to be statistically
significant.

Figure~\ref{fig:lumfuncz} shows how the comoving luminosity function
of these groups evolves with redshift.  Panel \emph{(a)} shows the
total \lya luminosity of each group, while the others include
only IGM emission.  Figures~\ref{fig:lumfuncz}\emph{a} and
\ref{fig:lumfuncz}\emph{b} assume that the self-shielded gas is in
CIE, Figure~\ref{fig:lumfuncz}\emph{c} assumes that all gas is
optically thin, and Figure~\ref{fig:lumfuncz}\emph{d} sets $\epsa=0$ for
self-shielded gas.  We use the Q5 simulation for $z=5$, the D5
simulation for $z=3$ and $2$, and the G5 simulation for $z=0$.  We
show sample error bars for the D5 simulation at $z=3$ assuming Poisson
statistics.  Note that the sharp cutoffs at large luminosities for the
$z \ge 2$ simulations are not real; they are due to the finite size of
the simulation boxes.  (However, the difference across panels in the
location of the cutoff is real.)  The turnover at small luminosity
for $z \le 3$ is due to the finite mass resolution and is also not real.

The trends with redshift are essentially as expected.  As emphasized
above, the total emission is dominated by SF, so panel \emph{(a)}
qualitatively resembles the ``SF multiplicity function'' of
\citet{springel03} (except that we have not divided low-redshift
groups and clusters into their constituent galaxies).  In particular,
${\rm d}n/{\rm d} \ln L$ declines at $z \la 1$ along with the global
SFR \citep{springel03}.  The IGM luminosity functions depend
principally on the treatment of self-shielded gas.  Groups at $z \sim
2$--$4$ are most luminous if we assume CIE, while those at $z \sim 0$
and $z\sim5$ are most luminous if all gas is optically thin.  As
before, this is a direct consequence of the temperature of the cooling
locus.  The maximum IGM luminosities without self-shielded gas are
much smaller. If this were the most accurate model then it would be
considerably more difficult to detect the coronae in \lya emission.

\begin{figure}[t]
\resizebox{9cm}{!}{\includegraphics{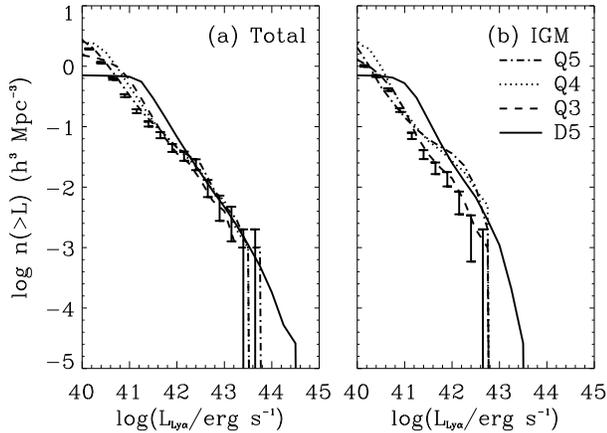}}
\caption{The Ly$\alpha$ luminosity function of groups at $z=3$ in the
  different simulations.  \emph{(a):} Total luminosity function,
  including both IGM (assuming CIE) and SF. \emph{(b):} Luminosity
  function for IGM emission only, assuming CIE.  All distances are
  comoving.  Error bars are Poissonian.}
\label{fig:lumfunc3}
\end{figure}

Figure~\ref{fig:lumfunc3} shows the luminosity function at $z=3$ in
several different simulations.  Here we show the \emph{cumulative}
function for comparison with existing observations (see \S \ref{obs}).
Panel \emph{(a)} again shows the total \lya luminosity of each group,
while panel \emph{(b)} includes only the IGM emission.  In both cases
we assume that the self-shielded gas is in CIE.  These show clearly
that our luminosity functions have converged over most of the
appropriate range and that the low and high $L_{\rm Ly\alpha}$
cutoffs are not physical.  We find similar behavior for the other
treatments of dense gas.

As in \S \ref{maps}, we find that the IGM emission depends sensitively
on our assumptions about dense gas: the luminosity function of this
component varies by several orders of magnitude in the different
cases.  Thus observations of (or constraints on) the diffuse emission
around galaxies will help to determine how gas accretes onto galaxies
and how this gas interacts with the ionizing background (diffuse and
local) and with galactic winds.  Like \citet{fardal01}, we find that
the \lya emission from SF typically dominates the total luminosity,
even in the most optimistic IGM treatments.  However, the disparity
between the two components is not as large as \citet{fardal01} claim
unless $\epsa=0$ for self-shielded gas (see their Figure~2).  One
possible explanation is that the conservative formulation of SPH
employed in our simulations is much less susceptible to numerical
overcooling problems that troubled earlier work \citep{springel-ent}.
Another is that \citet{fardal01} used a simulation without a
photoionizing background to compute the IGM emission.  We have argued
that this is not a good assumption in general, although it may be
adequate if the gas becomes self-shielded before most of the
gravitational heating.  In that limit their method is in principle a
cleaner test of the importance of gravitational shock-heating.  Their
results lay between our three cases, suggesting that (1) a substantial
fraction of the gravitational energy is radiated \emph{after} the gas
becomes self-shielded and/or (2) more of the gravitational energy is
radiated through other cooling mechanisms in our simulations.  The
former is not inconsistent with our results, because the self-shielded
gas has contracted more than the optically thin component, which lies
in the outskirts of the coronae.  We also note that SF emission is
much more likely to suffer from dust extinction, which would make the
IGM emission easier to see along those lines of sight.

\subsection{Characteristic Sizes}
\label{sizes}

We use SExtractor (version 2.3.2, \citealt{bertin96}) to identify
sources within our maps.  We choose the default parameter values with the
following exceptions: (1) we set the deblending parameter to 0.05 to
match the choice of M04, (2) we do not convolve the
image with a filter before identifying sources, and (3) we identify
sources using an absolute surface brightness rather than a multiple of
the local background.  We make this choice because our maps are (by
construction) background-free and we wish to avoid any artifacts from
imposing one.

To give some intuition, the green circles in Figure~\ref{fig:sez3map}
outline sources identified by SExtractor with a threshold of $\Phi =
100 \photunits$ (or $9.6 \times 10^{-21}\sberg$).  In each case, the
area of the circle is proportional to the isophotal area of the object
(though we do not outline the shape of the individual sources).  We
have set the minimum area to 2 square arcseconds for completeness
here.  The largest source in the left slice has $A=237$ square
arcseconds for SF.  It splits into two sources for the IGM,
with $A=185$ and 182 square arcseconds.  Only five (or three) other
sources have $A>10$ square arcseconds for IGM (or SF).  In
the right-hand map, only two sources have $A>10$ square arcseconds, in
both IGM and SF.  In general, the IGM \lya emission is
more spatially extended but fainter than the associated emission from
SF.

\begin{figure}[t]
\plotone{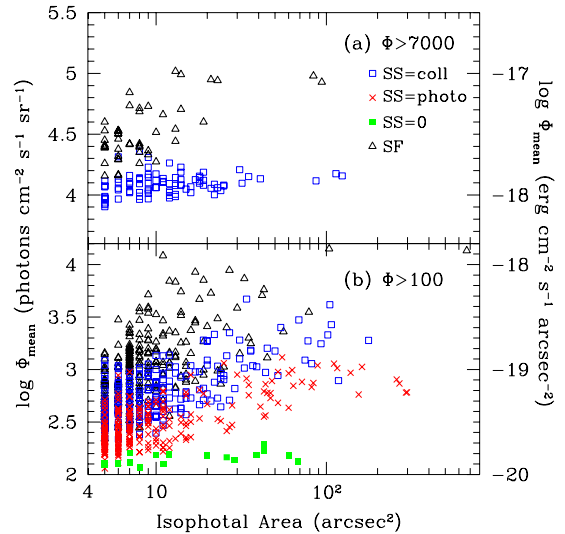}
\caption{Scatter plot of angular size against mean surface brightness
  for sources from the Q5 simulation at $z=3$.  Panels \emph{(a)} and
  \emph{(b)} have $\Phi=7000$ and $100 \photunits$ (or $6.7 \times
  10^{-19}$ and $9.6 \times 10^{-21}\sberg$), respectively.  The
  triangles are for emission from SF; other symbols are
  for IGM emission with the different treatments of self-shielded gas.
  Note that $1\arcsec = 21.6 h^{-1}$ comoving kpc. }
\label{fig:size}
\end{figure}

Figure~\ref{fig:size} shows a scatter plot of isophotal area versus
mean surface brightness for slices from the Q5 simulation at $z=3$.
We have used ten slices of thickness $\Delta z=0.01$ and width $5.5
h^{-1} \Mpc$ (comoving) to construct the sample.  The angular
resolution is $1\arcsec$ (equal to $21.6h^{-1}$ comoving kpc).  The
total volume is thus about twice the volume of the Q5 box (but most
repeated sources will be viewed from different directions).  In all
cases we have imposed a minimum source area of 5 square arcseconds.
The triangles are for SF, while the other symbols show the IGM
emission for our three cases.  Figures~\ref{fig:size}\emph{a} and
\emph{b} have detection thresholds of $\Phi = 7000 \photunits$ (or
$6.7 \times 10^{-19} \sberg$) and $\Phi=100 \photunits$ (or $9.6
\times 10^{-21} \sberg$), respectively.  For the former,
there are no IGM emitters that meet this threshold unless the
self-shielded gas is in CIE; the other samples are complete.  For the
lower threshold of Figure~\ref{fig:size}\emph{b}, we randomly select
250 objects from the samples (unless $\epsa=0$ for self-shielded gas,
in which case only 16 sources meet our criteria).
 
Comparing the IGM and SF points at the two $\Phi$ thresholds, we see
that the IGM emission is usually fainter but more extended than the SF
component.  This remains true until the highest thresholds, when the
IGM emission cuts off before SF emission; of course the location of
the cutoff depends on the treatment of dense gas.  At the flux levels
currently accessible to observations (Figure~\ref{fig:size}\emph{a})
IGM emission will be difficult to detect except where it is
self-shielded, has a high temperature, and is in CIE.  Interestingly,
in this case there are about twice as many IGM sources with $\Phi >
7000 \photunits$ as there are SF sources; this is because most of
the star-forming regions have $A<5$ square arcseconds.

\begin{figure}[t]
\plotone{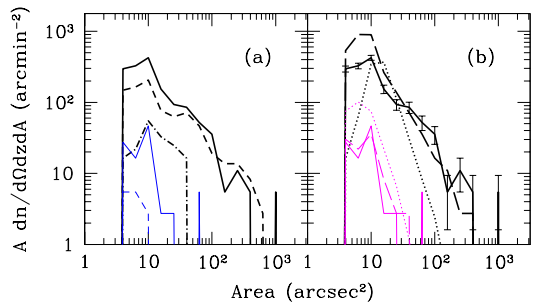}
\caption{ The distribution of Ly$\alpha$ emitters as a function of
  isophotal area at $z=3$.  The angular resolution is $1\arcsec$
  ($21.6 h^{-1}$ comoving kpc). \emph{(a):} The solid curves show
  total emission, while the dashed and dot-dashed curves show emission
  from the IGM only, assuming CIE for self-shielded gas and that all
  gas is optically thin, respectively.  Thick and thin curves set the
  threshold at $\Phi=10^3$ and $2.3 \times 10^4 \photunits$ (or $9.6
  \times 10^{-20}$ and $2.2 \times 10^{-18} \sberg$). All curves are
  from the Q5 simulation.  \emph{(b):} The solid, dashed, and dotted
  curves are for total (IGM and star formation) emission in the Q5,
  D5, and G6 simulations.  The cutoff at small area is due to the
  SExtractor threshold that we chose.  }
\label{fig:areafcn}
\end{figure}

\begin{figure}[t]
\plotone{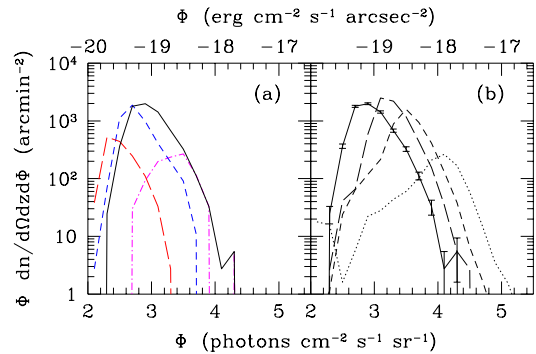}
\caption{ The distribution of Ly$\alpha$ emitters as a function of
  mean surface brightness at $z=3$.  The angular resolution is
  $1\arcsec$ ($21.6 h^{-1}$ comoving kpc). \emph{(a):} The solid,
  short-dashed, and long-dashed lines are for the total emission, IGM
  assuming CIE for self-shielded gas, and IGM assuming that all gas is
  optically thin; each of these has $A_{\rm min}=5$ square arcseconds.
  The dot-dashed line shows the total emission with $A_{\rm min}=16$
  square arcseconds.  \emph{(b):} The solid, long-dashed,
  short-dashed, and dotted curves are for total emission in the Q5,
  Q3, D5, and G6 simulations, with $A_{\rm min}=5$ square arcseconds.}
\label{fig:phifcn}
\end{figure}

Figures~\ref{fig:areafcn} and \ref{fig:phifcn} show the source
characteristics in a more quantitative form.  Figure~\ref{fig:areafcn}
shows the angular density of \lya sources per unit redshift as a
function of isophotal area (note that the cutoff at small area is a
result of the minimum detection area we set in SExtractor).  Panel
\emph{(a)} compares the total (including SF and assuming CIE for
self-shielded gas, solid curve) and IGM (dashed curves for CIE,
dot-dashed curve for optically thin gas) emission at two different
flux thresholds, $\Phi=10^3$ and $2.3 \times 10^4 \photunits$ (or $9.6
\times 10^{-20}$ and $2.2 \times 10^{-18} \sberg$; thick and thin
curves, respectively).  Note that there are no objects with
$\Phi > 2.3 \times 10^4 \photunits$ if all gas is optically thin.
Again, we find that the isophotal area of the total emission is
dominated by the IGM at small $\Phi$, but at sufficiently large
surface brightness thresholds, the SF component dominates by a large
factor.  

Figure~\ref{fig:areafcn}\emph{b} shows the total emission for the Q5
(solid), D5 (dashed), and G6 (dotted) simulations.  We also show error
bars on the Q5 curve, assuming Poisson statistics.  Clearly the G6
simulation underestimates both the number of faint, compact sources
and the number of highly extended sources.  The former occurs because
the simulations have a finite mass resolution.  The latter probably
occurs because the larger-scale simulations do not fully resolve the
spatial distribution of SF within halos, even though the global SFR
converges (see \S \ref{simres}).  The higher-resolution simulations do
seem to have converged reasonably well.

Figure~\ref{fig:phifcn} shows similar information, except as a
function of the mean surface brightness.  Panel \emph{(a)} shows that
the SF component dominates the total brightness, especially if the
dense gas is optically thin (or if it cannot emit).  Note also that a
larger area threshold decreases the abundance of both faint and
extremely bright sources.  Figure~\ref{fig:phifcn}\emph{b} shows poor
convergence in the brightness distribution: the number of sources with
$\Phi > 10^4 \photunits$ varies by more than an order of magnitude
between the simulations.  This may be surprising given that the \lya
luminosity function converges reasonably well.  It is again a result
of failing to resolve the spatial distribution of the gas.  The
high-resolution simulations better describe the extended, cool gas, so
the sources have larger areas and hence smaller mean $\Phi$ (even for
the same intrinsic luminosity).

We note that winds have little effect on the sizes of the \lya
emitters.  There is weak evidence that the low surface brightness IGM
emission is slightly more extended in the case of strong winds, but it
is not a significant difference.  This is not surprising given that
the winds will have difficulty escaping along the direction of bright,
dense particles.

In this section we have focused on $z=3$, because that is where the
most complete observations are.  However, our qualitative conclusions
hold at other redshifts as well:  star-forming regions produce
compact, bright \lya emission surrounded by extended, fainter IGM
emission.  

\subsection{Comparison to Observations}
\label{obs}

In the past several years, \lya line selection has become an
increasingly popular technique to identify high-redshift objects.  The
usual procedure is to compare broadband and narrowband images of a
field; those objects that have high equivalent widths in the
narrowband image are likely to be emission-line galaxies.  In most
cases, contamination from other lines (particularly [\ion{O}{2}]
$\lambda3727$ at low redshifts) is a worry, but color selection and
spectroscopic follow-up can determine the sample purity.  To date, most
studies have focused on relatively high-surface brightness emission
from star-forming galaxies.  

A number of these surveys have constrained the space density of
luminous, relatively compact \lya emitters.  We will begin by
comparing these observations to our results.  \citet{cowie98} performed
blank-field surveys at $z\approx3.4$ and found $n \sim (3,1) \times
10^{-3} h^3 \Mpcden$ for objects with $L_{42} \ga (2,5)$.  Here
$L_{\rm Ly\alpha} = L_{42} \times 10^{42} \ergs$.  This is about an
order of magnitude smaller than the predictions of
Figure~\ref{fig:lumfunc3}.  (Note that the luminosity function evolves
only weakly with redshift in the range $z \sim 2$--$5$.)
S00 found a space density about 6 times larger than
\citet{cowie98}; however, the S00 volume contains $\sim
6$ times as many continuum-selected galaxies as an average field, so
this is probably due to cosmic variance.  \citet{fujita03} find $n
\sim 2 \times 10^{-4} h^3 \Mpcden$ for objects with $L_{42} \ga 5$ at
$z=3.7$, several times smaller than \citet{cowie98}.  \citet{ouchi03}
performed a survey for \lya emitters at $z=4.86$ and found no evidence
for evolution in the luminosity function between that time and
$z=3.4$. \citet{palunas04} found $n \sim 3 \times 10^{-4} h^3 \Mpcden$
for $L_{42} \ga 4$ at $z=2.38$, also consistent with weak redshift
evolution.

All of these surveys have different selection criteria, so detailed
comparisons to our luminosity function are difficult.  However, it
seems clear that our predictions are roughly an order of magnitude
above the observed number densities at $L_{42} \sim 1$--$10$.  There
are several possible explanations for the discrepancy.  First, we have
included the total emission from SF and the IGM.  The IGM emission is
relatively extended and may not actually contribute to the observed
high-surface brightness emission.  If the CIE case is overly
optimistic, this would also decrease the luminosities by about $50\%$.
Second, we have ignored processes within the galaxies that can
dramatically modify the \lya line luminosity, such as dust (which
appears to completely eliminate or substantially reduce \lya emission
in $\sim 75$--$80\%$ of $z=3$ galaxies; S00, \citealt{shapley03}).
Such a correction would bring our estimates into reasonable agreement
with most of the observations described above.  The dust extinction
probably depends on the line of sight through the galaxy (and its
wind); much of the galaxy will therefore show little \lya emission,
allowing the IGM to stand out more clearly.  Third, we have used a
crude conversion of SFR to $L_{\rm Ly\alpha}$ (see \S \ref{sflum}).  A
more sophisticated treatment may help to reconcile the predictions
with the observed number density.

To probe the IGM emission, it is best to consider extended objects
with relatively low surface brightness.  S00 and M04 have detected
extended ``blobs'' of \lya emission without associated (significant)
UV continua near a ``proto-cluster'' of Lyman-break galaxies.  S00
detected two blobs with $L_{\rm Ly\alpha} \sim 10^{44} \ergs$ and
physical extents of $\sim 100 h^{-1} \kpc$, implying $n \sim 3 \times
10^{-4} h^3 \Mpcden$ for these objects.  (Note, however, that this is
the same field cited above and has $\sim 6$ times as many galaxies as
average).  M04 surveyed an overlapping field (with effective volume
about eight times larger) and selected objects with $A > 16$ square
arcseconds and $\Phi \ga 2.3 \times 10^4 \photunits$ ($2.2 \times
10^{-18} \sberg$); they find $n \sim 7 \times 10^{-4} h^3 \Mpcden$.
These objects have $L_{42} \ga 6$.

Extended emission is thus somewhat rarer than the compact \lya
emitters probed by most surveys, and our models can easily accommodate
the raw luminosities of the M04 sample.
Figure~\ref{fig:lumfunc3}\emph{b} shows that the IGM component on its
own can provide enough energy so long as we assume CIE.  However, most
of the sources in the simulation are much smaller than the M04
objects.  With their area and surface brightness thresholds, we find
$n_{\rm sim} \sim (3,20) \times 10^{-4} h^{3} \Mpcden$ for IGM and
total emission, respectively, assuming CIE for self-shielded gas.  In
the other cases \emph{no} IGM objects satisfy these criteria.
Unfortunately, these predictions are based on only a few objects and
have not fully converged for the IGM emission.  The strongest
statement we can make is that star formation appears to produce easily
enough \lya emission on moderately large ($A \sim 20$--$40$ square
arcsecond) scales to explain the observations, even if we renormalize
the number densities to match that of \lya galaxies (thus crudely
accounting for dust extinction in the SF component).  However, the IGM
on its own could only account for the blobs in the most optimistic
case.

If we consider the most luminous and most extended objects, the IGM
looks to be even less plausible.  At the M04 $\Phi$ threshold,
\emph{none} of our objects have $A > 60$ square arcseconds even if we
include star formation, while $\sim 10\%$ of the observed blobs exceed
this size (and the largest has $A=222$ square arcseconds).  We most
likely therefore need to appeal to other processes to explain the
largest and most luminous blobs (including those first discovered by
S00), although we cannot rule out that the finite size of our
simulation box is responsible for the difference.  We must also stress
that the observed luminous \lya blobs have velocity widths of several
hundred km s$^{-1}$ (S00; \citealt{bower04}), much larger than the
intrinsic widths of our sources (see \S \ref{vel}).  If these sources
are ultimately due to cooling gas and SF, large internal velocity
gradients (from winds, for example) or complex radiative transfer
(e.g., \citealt{zheng02}) would have to dramatically alter our
predicted linewidths.  On the other hand, both S00 and M04 selected a
region known to contain a strong galaxy overdensity, and the density
of blobs in more normal environments is unknown.  M04 note that the
blobs are highly clustered even within their region.  Observations of
more representative regions would be invaluable in understanding the
nature of strong \lya emission.

Another interesting issue is the correlation between IGM emitters and
galaxies.  In our simulations, \lya sources without associated stellar
particles are extremely rare.  In most cases, those that do exist are
associated with poorly-resolved halos and may be simulation artifacts.
Thus, within the context of our simulations, isolated bright \lya
emitters will be rare.  Although the luminous S00 blobs are not
centered on optically visible galaxies, at least one of them has a
vigorously star-forming (SFR $\ga 500 h^{-2} \sfr$) submm galaxy near
its center \citep{chapman01} and both have optical galaxies near their
outskirts.  M04 found that 7/33 other blobs match with previously
detected galaxies.  The others may also be associated with faint
galaxies (perhaps due to intrinsic absorption), and testing this
connection would be useful in the future.  Such observations could
also help to measure the relative importance of \lya from SF and the
IGM in different galaxies.  We have found that SF usually dominates
the total luminosity of each system; however, we have also found that
our simulations overestimate the number density of \lya-emitting
galaxies.  This is likely due to internal structure and the dust
distribution within galaxies.  The IGM emission should be less
susceptible to these uncertainties, so in some cases it may actually
dominate the observed emission.

Finally, we have neglected radiative transfer in computing the
source sizes.  Because our regions are optically thick in the \lya
line, these photons scatter resonantly until reaching an optical
depth near unity.  Although most of the scattering occurs in frequency
space, some photons could in principle also scatter in real space to
the edge of the optically thick region, from which they can escape
even from the center of the line.  Thus our method may underestimate
the true extent of the emitting regions.  \citet{fardal01} argue that
this mechanism can help to reconcile the luminous, compact sources
found in their simulations with the significantly more extended
observed systems.  Furthermore, ionizing photons from young stars in
the central galaxy can escape into the surrounding IGM but get
absorbed by the cool, dense gas.  This would shift some of the \lya
emission that we attribute to star formation into the IGM, making the
SF sources larger than we have found.

\section{Discussion}
\label{disc}

We have used cosmological simulations to study the \lya emission
expected from structure formation over a broad range of redshifts.  We
considered emission from optically thin low-density gas, from
self-shielded gas clouds, and from the reprocessing of stellar
ionizing photons.  As in F03, who studied \lya emission from $z \la
0.5$, we have found that the brightest regions are compact
star-forming galaxies.  These are surrounded by extended coronae of
emission from IGM gas; the regions closest to galaxies are likely to
be self-shielded (at least from the mean ionizing background) while
the outer reaches remain optically thin.  Crucially, although emission
associated with star formation dominates the total \lya luminosity,
the IGM coronae are more extended and can thus be isolated (at least
in principle).  They could also dominate in galaxies in which dust has
extinguished the \lya line from star formation.  However, the
luminosities of the coronae depend sensitively on our assumptions
about gas that is able to shield itself from the metagalactic ionizing
background.

One of the most important results of this study is that the
photoionizing background can power the bright \lya coronae, in
contrast to earlier predictions in which gravitational energy
dominated \citep{haiman00-lya,fardal01}.  In our simulations, most of
the flux comes from gas on a high density ($\nh \ga 10^{-3} \cmden$),
low temperature ($T \la 10^{4.5} \kel$) track of particles bound to
dark matter halos but outside of galaxies.  Although many of these
dense particles have been moderately shocked in the past (see
Figure~\ref{fig:trace}), the post-shock cooling times are so short
that they quickly enter a quasi-equilibrium cool phase.  In this
regime, photoheating and radiative cooling nearly balance, increasing
the cooling time.  While it is optically thin, the ionizing background
powers $\ga 50\%$ of the \lya emission at $z \ge 1$, provided that it
is near or above our fiducial amplitude.  As the gas condenses, we
expect it to eventually become self-shielded from the diffuse ionizing
background.  Because we cannot include radiative transfer in our
calculations, the power source beyond this point is unclear.  If
gravitational contraction or winds maintain the gas temperature near
the peak for \lya excitation, photoionization would play a relatively
minor role.  If, on the other hand, the dense gas rapidly cools or
remains optically thin (in particular due to embedded sources),
photoionization can power $\ga 90\%$ of the \lya emissivity.  We thus
argue that including photoionizing radiation is crucial to
interpreting observations of \lya cooling around galaxies.

In any of these cases, the cool gas in \lya coronae directly traces
the accretion of gas onto galaxies, although winds do significantly
complicate the interpretation.  In the presence of winds, \lya
emission comes from a mix of relatively pristine accreting gas and
material that has recently been ejected from the central galaxy but
remains trapped inside the halo (either because the wind velocity is
smaller than the escape speed of the halo, or because the wind
interacts with the accreting gas).  We find, however, that winds have
a surprisingly small effect on the \lya emission (especially
considering that they reduce the global SFR by a factor of five),
although our simulations do not fully resolve the winds.  Clumping
within the wind medium could increase the local emissivity, or the
faster cooling may end up reducing the total emission.

Thus the \lya coronae present the intriguing opportunity to measure
some combination of wind feedback near and gas accretion onto
galaxies.  Unfortunately, the self-shielded gas responsible for most
of the emission is also the most difficult to model.  We have
emphasized several uncertainties in \S \ref{ss}, including the
presence of dust, the geometry and velocity structure of the region,
radiative transfer of higher Lyman line photons through the cloud, the
temperature of the self-shielded gas, and the local ionizing radiation
field.  Of these, the last two are likely the most important.  

The temperature of gas on the cooling locus is determined in the
simulation by the balance between photoheating and radiative cooling.
If the cloud is self-shielded, the primary heat source disappears and it
may cool rapidly.  The \lya emissivity is extremely sensitive to the
temperature in collisionally ionized gas, so small errors in the heat
balance could translate into large changes in the emission properties.
Unfortunately, the simulations do not include self-shielding because
they do not incorporate a self-consistent treatment of radiative
transfer.  Although we can crudely identify regions that are shielded
from the metagalactic ionizing background, we cannot account for
radiation from nearby galaxies or self-consistently describe the
behavior of shielded gas.  We therefore consider three cases that
should bracket the real behavior.  In the first, we set the emissivity
of self-shielded gas to zero, which would be a good approximation if
the gas were to cool efficiently.  In the second, we assume that the
simulation temperatures are accurate and that the self-shielded gas is
in CIE.  In the third, we assume that the local ionizing field is much
larger than the mean, so that even dense gas remains optically thin.
In the first case, IGM gas is extremely faint and may be beyond the
reach of near-future surveys.  But in either of the other cases, the
gas is significantly brighter and the total luminosity is not too far
below that from star formation.  The balance between the two
optimistic cases depends on redshift, because the emissivity in CIE is
extremely sensitive to temperature, which, in our simulations, is set
by the ionizing background.

We have also computed the \lya emission from recombinations following
absorption of stellar ionizing photons inside galaxies.  We have used
an extremely simple conversion between local SFR and \lya luminosity
(see \S \ref{sflum}), appropriate for a standard IMF and little
extinction.  The true luminosity will vary with the initial mass
function, metallicity, dust content, and internal structure of each
galaxy, so our quantitative results should be viewed as no more than
representative.  With these assumptions, this component dominates both
the total luminosity and the brightest pixels, but it is confined to
smaller physical scales than the IGM emission.  We have, however,
assumed that all of the ionizing photons are absorbed locally.  In
reality, some fraction of the ionizing photons will escape the host
galaxy into the surrounding IGM \citep{steidel,lequeux,kunth03}.  It
is this radiation that could escape and help to illuminate the
nominally self-shielded gas in the \lya coronae.  By neglecting dust,
we have also overestimated the total amount of SF \lya radiation by a
factor of $\sim 4$ (at $z=3$; S00).  Both of these effects help to
increase the relative importance of IGM emission.

A number of recent observations have targeted high-redshift galaxies
with strong \lya emission (see \S \ref{obs}).  The space density of
such objects is about an order of magnitude smaller than predicted by
our simulations.  A large part of this difference is likely because
only $\sim 20$--$25\%$ of star-forming galaxies have strong \lya
emission (S00), for the reasons outlined above.  The remaining
discrepancy is reasonable given the simple assumptions behind our
model.  Our results therefore suggest that \lya selection can be a
useful probe of galaxy formation and gas accretion, even if
high-surface brightness galaxies are the only targets.  This is
particularly interesting given that \lya-selection is currently the
most efficient technique to select the highest-redshift galaxies
(e.g., \citealt{hue02,kodaira03,rhoads04}) and can potentially be a
powerful probe at even earlier epochs (\citealt{barton04}; though
absorption from the neutral intervening IGM could complicate such an
interpretation; see, e.g., \citealt{furl04d} and references therein).

We have emphasized the substantial uncertainties in \lya emission from
both the IGM and star formation.  While these imply that our
predictions are far from rigorous, they present an excellent
opportunity to constrain the models through observations.  Existing
surveys for diffuse \lya emission at $z\sim2$--$3$ have reached
surface brightness thresholds of $\sim 2.3 \times 10^4 \photunits$ (or
$2.2 \times 10^{-18} \sberg$ at $z=3.1$; M04) with the Subaru
telescope and $\sim 1.3 \times 10^4 \photunits$ (or $1.6 \times
10^{-18} \sberg$ at $z=2.2$; \citealt{francis04}) with the 4-meter
Anglo-Australian Telescope.  Both are fairly close to our more
optimistic predictions.  We have shown that the combination of star
formation and cooling gas powered by the photoionizing background can
in principle account for most of the \lya blobs observed by M04,
although we found no counterparts to the largest and most luminous
blobs identified by S00.  IGM emission alone probably cannot explain
the observations, but in our more optimistic cases should appear with
order-of-magnitude increases in the surface brightness sensitivity.

We find no evidence in our simulations for strong \lya emission except
that surrounding galaxies.  We therefore expect that all the \lya
blobs observed at $z \sim 3$ contain embedded galaxies, which could be
tested through follow-up observations of the M04 sources.  An
alternate approach to find isolated emitters is to seek quasar
absorption line systems in emission \citep{hogan87}.  Analytic
estimates suggest that optically-thick Lyman-limit systems should be
visible through reprocessing of the ionizing background
\citep{gould96}.  In our simulations, it is most natural to associate
these with the \lya coronae around galaxies, so it is important to
identify the absorber environments through observations.  Such
searches are already underway.  \citet{francis04} searched for diffuse
emission near a bright quasar at $z=2.168$.  They found no sources
with $A>9$ square arcseconds and $\Phi > 1.3 \times 10^4 \photunits$.
This is somewhat surprising in our model, given that we expect a
quasar to be embedded inside an overdensity with an abundance of cool
gas; the peak brightness should be at least comparable to their
threshold if the gas emits in collisional ionization equilibrium.  The
most likely explanation is that the quasar has photoionized all of the
nominally self-shielded gas.  This would reduce the peak to $\Phi \sim
10^3$-$10^4 \photunits$ (see Figure~\ref{fig:zevol}\emph{c}), just
below the existing threshold.  We therefore argue that it is crucial
to continue to push these narrowband observations forward.
\citet{francis04} have demonstrated that valuable upper limits can be
obtained even from 4-meter class telescopes, and our results suggest
that direct detection of diffuse \lya emission may not be far away.

\acknowledgements We thank G. Becker, W. Sargent, and C. Steidel for
enlightening discussions.  We also thank G. Ferland for invaluable
help in interpreting the Cloudy results.  SRF thanks the Institute of
Advanced Study, where part of this work was completed.  This work was
supported in part by the W.~M.~Keck foundation and NSF grants ACI
96-19019, AST 00-71019, PHY-0070928, AST 02-06299, and AST 03-07690,
and NASA ATP grants NAG5-12140, NAG5-13292, and NAG5-13381.  The
simulations were performed at the Center for Parallel Astrophysical
Computing at Harvard-Smithsonian Center for Astrophysics.

\end{document}